\documentstyle[preprint,aps,eqsecnum]{revtex}
\def\tr{{\rm Tr}}

\newcommand{\be}{\begin{eqnarray}}
\newcommand{\ee}{\end{eqnarray}}
\newcommand{\BE} {\begin{equation}}
\newcommand{\EE} {\end{equation}}

\begin{document}
\draft
\tighten

\title{
\Large\bf A Weak-Coupling Treatment of Nonperturbative QCD
	Dynamics to Heavy Hadrons}

\author{{\bf Wei-Min Zhang} \\
	Institute of Physics, Academia Sinica, Taipei 11529, Taiwan\thanks{
	The current address, E-mail address is: wzhang@phys.sinica.edu.tw} \\
	Department of Physics, National Tsing Hua University 
	  Hsinchu 30043, Taiwan }

\date{Sept. 10, 1996, Revised April 10, 1997}

\maketitle

\begin{abstract}
Based on the recently developed light-front similarity renormalization 
group approach and the light-front heavy quark effective theory, we
derive analytically from QCD a heavy quark light-front Hamiltonian which
contains explicitly a confinement interaction at long distances and  
a Coulomb-type interaction at short distances. With this light-front 
QCD Hamiltonian, we further demonstrate that the nonperturbative QCD 
dynamics of the strongly interacting heavy hadron bound states can be
treated as a weak-coupling problem.
\end{abstract}

\vspace{0.3in}


\newpage

\section{Introduction}

There are two fundamental features in nonperturbative QCD,
the quark confinement and the spontaneous breaking of chiral 
symmetry. These two features are the basis for solving 
hadronic bound states from QCD but none of them has been completely 
understood.  Recently, Wilson et al. proposed a new approach to 
determine hadronic bound states from nonperturbative QCD on the 
light-front with a weak-coupling treatment (WCT) \cite{Wilson94}. 
The key to eliminating necessarily nonperturbative effects is to 
construct an effective QCD Hamiltonian in which quarks 
and gluons have nonzero constituent masses rather than the 
zero masses of the current picture.  The use of constituent
masses cuts off the growth of the running coupling constant
and makes it conceivable that the running coupling never
leaves the perturbative domain. The WCT approach 
potentially reconciles the simplicity of the constituent quark
model with the complexities of QCD. The penalty for achieving 
this weak-coupling picture is the necessity of formulating 
the problem in light-front coordinates and of dealing with 
the complexities of renormalization. 

Succinctly, this new approach of achieving a QCD description of 
hadronic bound states can be summarized as follows\cite{Wilson94}: 
Using a new renormalization scheme, called similarity renormalization 
group (SRG) scheme \cite{Glazek94}, one can obtain an effective
QCD Hamiltonian $H_\lambda$ which is a series of expansion in 
terms of the QCD coupling constant, where $\lambda$ is a low 
energy scale ($\simeq 1$ GeV). Then one may solve from $H_\lambda$  the 
strongly interacting bound states as a weak-coupling problem.
The WCT scheme contains the following steps: (i) Compute 
explicitly from SRG the $H_\lambda$ up to the second order 
and denote it by $H_{\lambda 0}$ as a nonperturbative part 
of $H_\lambda$.  The remaining higher order contributions in 
$H_\lambda$ are considered as a perturbative part $H_{\lambda I}$. 
(ii) Introduce a constituent picture which allows one to start 
the hadronic bound states with the valence constituent Fock space.  
 The constituent quarks and gluons have masses of 
a few hundreds MeV, and these masses are functions of the scale
$\lambda$ that must vanish when the effective theory goes back 
to the high energy region. (iii) Solve hadronic bound states 
with $H_{\lambda 0}$ nonperturbatively in the constituent picture and 
determine the scale dependence of the constituent masses and the 
coupling constant. The coupling constant $g$ now becomes an effective 
one, denoted by $g_\lambda$. If we could show that with a suitable 
choice of $\lambda$ at the hadronic mass scale, the effective
coupling constant $g_\lambda$ can be arbitrarily small, then 
WCT could be applied to $H_\lambda$ such that the corrections 
from $H_{\lambda I}$ can be truly computed perturbatively. 
If everything listed above works well, we may arrive at a 
weak-coupling QCD theory of the strong interaction for 
hadronic bound states. 

With the idea of SRG and the concept of coupling coherence
\cite{Perry93}, Perry has shown that upon a calculation to the 
second order, there exists a logarithmic confining potential in 
the resulting light-front QCD effective Hamiltonian \cite{Perry94}. 
This is a crucial finding to light-front nonperturbative QCD.  
However, the general strategy of solving hadrons through the WCT 
scheme has not been examined.  In this paper, I will use SRG to 
analytically derive from the light-front heavy quark effective theory 
(HQET) \cite{Zhang95,Cheung95} a heavy quark QCD Hamiltonian 
which is responsible to heavy hadron bound states. The resulting 
Hamiltonian explicitly contains a confining interaction between 
heavy quark and antiquark at long distance plus a Coulomb-type 
interaction at short distance. Based on this effective QCD Hamiltonian, 
I then study the strongly interacting heavy hadronic bound states,
from which I can qualitatively provide a realization of WCT 
to nonperturbative QCD on the light-front.

The reason of choosing heavy hadron systems as a
starting example in this investigation is that for light quark 
systems, both  quark confinement and spontaneously
chiral symmetry breaking play an essential role to the quark 
dynamics in hadrons. In other words, to provide a good QCD
description for light quark systems, it is necessary to 
understand the underlying mechanism for quark confinement 
as well as for chiral symmetry breaking. This will certainly 
complicate the study of nonperturbative QCD. However,  
for heavy quark systems, chiral symmetry is explicitly 
broken so that confinement is the sole nontrivial 
feature influencing heavy quark dynamics. 
One may argue that the mass scales for heavy and light 
quark systems are different. The heavy quark energy cannot 
run down to the usual hadronic scale of light quark systems due 
to heavy quark mass.  Meanwhile, confining interactions must be 
energy scale dependent.  Apparently, confinement for heavy quark 
systems could be very different from light quark systems.
However, despite heavy and light quarks, confinement arises only
from low energy gluon interactions.  In other words, the 
confinement mechanism must be the same for both heavy and 
light quark systems.  We thus choose heavy hadron systems  
without any loss of generality.  

In order to avoid the possible confusion arisen from different 
mass scales and to correctly extract confining interactions 
in heavy quark dynamics, it is convenient to work with 
heavy quark effective theory (HQET).  HQET 
is a theory of QCD in $1/m_Q$ expansion \cite{Georgi}, where 
$m_Q$ is the heavy quark mass. In HQET,
the nonperturbative dynamics is determined through 
the interacting gluons and heavy quarks by exchanging a small 
residual momentum of heavy quarks, which 
is of order $\Lambda_{QCD}$. As a result, within HQET we can 
indeed explore the nonperturbative QCD dynamics for heavy quark systems
in the same scale as that for light quark systems. Meanwhile, the 
extension of the study to light quark systems is straightforward, 
although undoubtedly the corresponding result must be very 
complicated due to the spin dependence of the nonperturbative
interacting Hamiltonian. The spin dependent interactions on the 
light-front are essentially related to the chiral symmetry breaking. 
These spin dependent interactions in HQET are suppressed
in the leading order approximation because they are  
$1/m_Q$ corrections and can be treated perturbatively 
with respect to heavy hadron states.  This is why 
for heavy quark systems the chiral symmetry 
breaking can be treated separately from the confinement.

In fact, the model-based theoretical investigations on heavy 
quarkonia lasted for one and half decades is recently replacing by 
first-principles exploration on QCD.  The lattice QCD simulation 
may give an acceptable description for heavy
quarkonium spectroscopy with manageable control over all 
the systematic errors \cite{Lepage91}. The development of 
nonrelativistic QCD provides a general factorization formula
to quarkonium annihilation and production processes so that 
a rigorous QCD analysis may become possible \cite{Lepage95}. 
Meanwhile, in the past five years considerable progress has 
been made for heavy hadrons containing one heavy quark, due 
mainly to the discovery of the heavy quark symmetry (HQS) 
\cite{Isgur90} and the development of HQET \cite{Georgi} from QCD.
The HQS and HQET have in certain contents put the description 
of heavy hadron physics on a QCD-related and model-independent 
basis. Moreover, HQET has also been extended to describe heavy 
quarkonia \cite{Mannel95}. Yet, a truly first-principles QCD 
understanding of heavy hadrons is still lacking because so far 
none is able to give a direct computation of heavy hadron bound 
states from QCD. On the other hand, in the last decade, the 
investigations of the light-front field theory on nonperturbative 
bound state problems have made some progress. Starting 
with heavy hadrons may provide a possible explicit solution 
of hadronic bound states in light-front QCD.  Hence, in this paper
we shall mainly concentrate on heavy quark systems, especially 
for heavy quarkonia, but the generality of nonperturbative
QCD dynamics obtained in this formalism will also be discussed.
 
The paper is organized as follows.  In Section 2, the general 
procedure of constructing a renormalized effective Hamiltonian 
$H_\lambda$ in the light-front SRG is briefly reviewed and the 
possible existence of light-front confining interactions in this
formulation is discussed. In Section 3, by applying
the general procedure to the light-front HQET of QCD, 
a light-front heavy quark confining Hamiltonian is
analytically derived.  A light-front picture of quark 
confinement is illustrated. In Section 4, 
the light-front heavy hadronic bound states are explored
within the WCT scheme.  As an example, the light-front
heavy quarkonium bound state equation is solved by the 
use of a Gaussian-type wavefunction ansatz, from which the WCT
to nonperturbative QCD is explicitly explored. Finally, the  
physical implications in the realization of WCT are discussed 
in Section 5.

\section{Effective QCD Hamiltonian in the SRG Scheme}

\subsection{Light-front SRG Scheme}

We begin with the general formulation of the similarity 
renormalization group approach to construct a low energy
QCD Hamiltonian which was first proposed by Glazek and Wilson
\cite{Glazek94}. The basic idea of the SRG approach is to develop 
a sequence of infinitesimal unitary transformations $S_\lambda$
that transform 
an initial bare Hamiltonian $H^B$ to an effective Hamiltonian
$H_\lambda$ in a band-diagonal form relative to an arbitrarily
chosen energy scale $\lambda$:
\begin{equation}
	H_\lambda = S_\lambda H^B S_\lambda^\dagger. \label{st1}
\end{equation}
Here the band-diagonal form means that the matrix elements of 
$H_\lambda$ involving energy jumps 
much larger than $\lambda$ will all be zero, while matrix elements
involving smaller jumps or two nearby energies remain in $H_\lambda$.
The similarity transformation should satisfy the condition that for
$\lambda \rightarrow \infty, H_\lambda \rightarrow H^B$ and $S_\lambda
\rightarrow 1$.

In this paper, we shall follow the formulation of SRG developed on 
the light-front\cite{Wilson94}. The effective Hamiltonian we seek is 
$H_\lambda$ with $\lambda$ being of order a hadronic mass ($\sim
1$ GeV). We begin with a given bare Hamiltonian which can be
written by $H^B = H_0 + H_I^B$, where $H_0$ is a bare free 
Hamiltonian and $E_i$ is its eigenvalue. Consider an 
infinitesimal transformation, then Eq.(\ref{st1}) is reduced to
\begin{equation} \label{st2}
	{d H_\lambda \over d \lambda} = [H_\lambda, T_\lambda]
\end{equation}
which is subject to the boundary condition $\lim_{\lambda\rightarrow
\infty} H_\lambda = H^B$, where $T_\lambda$ is a generator of the
similarity transformation.

To force the  Hamiltonian $H_\lambda$
becoming a band-diagonal form in energy space, we need to specify
the action of $T_\lambda$. This can be done
by introducing the scale $\lambda$ with $x_{\lambda ij} = {E_j - E_i
\over E_i + E_j + \lambda}$ into a smearing function $f_{\lambda ij} 
= f(x_{\lambda ij})$ such that
\begin{equation}
	f(x) = \left\{ \begin{array}{cc} 1 & x \leq 1/3 \\
		{\rm smoothly ~from~ 1~to ~ 0} & ~~1/3 < x < 2/3 \\
		0 & x \geq 2/3 \end{array} \right. , 	\label{sfd}
\end{equation}
and reexpressing Eq.(\ref{st2}) by
\begin{eqnarray}
	&& {d H_{\lambda ij} \over d\lambda} = f_{\lambda ij}
	  [H_{I \lambda}, T_\lambda]_{ij} + {d \over d\lambda}
	  (\ln f_{\lambda ij}) H_{\lambda ij}, \nonumber \\
	&& T_{\lambda ij} = {1 \over E_j - E_i} \Bigg\{(1-f_{\lambda ij})
	  [H_{I \lambda}, T_\lambda]_{ij} - {d \over d\lambda}
	  (\ln f_{\lambda ij}) H_{\lambda ij} \Bigg\}.
\end{eqnarray}
Here we have written $H_\lambda= H_0 + H_{I \lambda}$ because $H_0$ 
is invariant under transformations. And we have also used the 
notation $A_{ij} = \langle i | A | j \rangle$, where $|i\rangle$ 
and $|j\rangle$ are eigenstates of $H_0$.
Since $f(x)$ vanishes when $x\geq 2/3$, one can see that 
$H_{\lambda ij}$ does indeed vanish in the far off-diagonal 
region. It also can be seen that $T_{\lambda ij}$ is zero
in the near-diagonal region. 
The solutions for $H_{I \lambda}$ and $T_\lambda$ are 
\begin{equation}	 \label{htes}
	H_{I\lambda} = H^B_{I \lambda} + {\underbrace{[H_{I \lambda'},
	 T_{\lambda'}]}}_R~, ~~~~~ T_\lambda = H^B_{I \lambda T}
	+ {\underbrace{[H_{I \lambda'}, T_{\lambda'}]}}_T,
\end{equation}
where $H^B_{I \lambda ij} = f_{\lambda ij} H^B_{I ij}$, $H^B_{I\lambda Tij} 
= -{1\over E_j-E_i} \left({d\, \over d\lambda} f_{\lambda ij}\right) 
H^B_{Iij} $, and 
\begin{eqnarray}
	{\underbrace{X_{\lambda^\prime ij}}}_R &=& - f_{\lambda ij}
		\int_\lambda^\infty d\lambda^\prime X_{\lambda^\prime ij}, 
		\label{2} \\
	{\underbrace{X_{\lambda^\prime ij}}}_T &=& - {1 \over E_j - E_i}
		\Big({d\, \over d\lambda} f_{\lambda ij} \Big)
		\int_\lambda^\infty d\lambda^\prime X_{\lambda^\prime ij}
		+ {1 \over E_j -E_i} (1 - f_{\lambda ij}) X_{\lambda ij}.
		\label{3}
\end{eqnarray}
From eq.(\ref{htes}), one obtains an iterated solution for $H_{\lambda}$,
\begin{eqnarray}
	H_{\lambda} &=& \Bigg( H_0 + H^B_{I\lambda} \Bigg) + 
		\Bigg( {\underbrace{[H^B_{I\lambda^\prime}, 
		H^B_{I\lambda^\prime T}]}}_R \Bigg) \nonumber \\
	 && + \Bigg( {\underbrace{[{\underbrace{[H^B_{I\lambda^{
		\prime\prime}}, H^B_{I\lambda^{\prime\prime} 
		T}]}}_{R^\prime} , H^B_{I\lambda^\prime T}]}}_R 
		+{\underbrace{[H^B_{I\lambda^\prime}, 
		{\underbrace{[H^B_{I\lambda^{\prime\prime}}, 
		H^B_{I\lambda^{\prime\prime} T}]}}_{T^\prime}]}}_R \Bigg)
		+ \dots 	\nonumber \\ 
	& = & H^{(0)}_{\lambda}+H^{(2)}_{\lambda}+H^{(3)}_{\lambda}+\dots .
		\label{eh1}
\end{eqnarray}
Thus, through SRG, we eliminate the interactions between the states 
well-separated in energy and generate the effective Hamiltonian of 
eq.(\ref{eh1}). The expansion of eq.(\ref{eh1}) in terms of the 
interaction coupling constant brings in order by order the full 
theory corrections to this band diagonal low energy Hamiltonian.

Practically, we do not really compute $H_\lambda$ to all the orders 
in the expansion of eq.(\ref{eh1}). In the WCT scheme, we consider
the first two terms in (\ref{eh1}) as the nonperturbative part of 
$H_\lambda$ in the determination of hadronic bound states. Thus, 
in the fist step of WCT, we shall only compute $H_\lambda$ up to 
the second order in coupling constant. We hope that the remaining 
part can be handled perturbatively, as we have mentioned in the 
introduction and will be discussed more later. Up to the 
second order, eq.(\ref{eh1}) can be further expressed explicitly by
\begin{eqnarray}	
	H_{\lambda ij} &=& f_{\lambda ij} \left\{ H_{ij}^B + \sum_k 
		H^B_{Iik} H^B_{Ikj} \Big[ \frac{g_{\lambda jik}}
		{\Delta E_{ik}} + \frac{g_{\lambda ijk}}{\Delta 
		E_{jk}} \Big] + \cdots \right\} .  \label{eh11}
\end{eqnarray}
The front factor in the above equation indicates that $H_\lambda$ 
only describes long distance interactions (with respect to the scale
$\lambda$) which is responsible to hadronic bound states.  The 
function $g_{\lambda ij}$ is given by
\begin{equation}
	g_{\lambda ijk} = \int_\lambda^\infty d\lambda'
		f_{\lambda' ik} {d \over d \lambda'}
		f_{\lambda' jk}. 		\label{sd1}
\end{equation}

More explicitly, the bare Hamiltonian $H^B$ input in the above formulation
can be obtained from the canonical Lagrangian with a high energy 
cutoff that removes the usual UV divergences.  For light-front QCD 
dynamics, the bare Hamiltonian in our consideration is the canonical 
light-front QCD Hamiltonian that can be either obtained from the 
canonical procedure in the light-front gauge \cite{Zhang93a,Zhang93b}  
or generated from the light-front power counting rules \cite{Wilson94}.  
Instead of the cutoff on the field operators which is introduced 
in ref.\cite{Wilson94}, we shall use in this paper a vertex cutoff
to every vertex in the bare Hamiltonian:  
\begin{equation}
	\theta(\Lambda^2/P^+ - |p_i^- - p_f^-|),   \label{ctf}
\end{equation} 
where $p_i^-$ and $p_j^-$ are the initial and final state light-front 
energies respectively between the vertex, $\Lambda$ is the UV cutoff 
parameter, and $P^+$ the total light-front longitudinal momentum of the 
system we are interested in.  The theta function is defined as:
$\theta (x) = 1, 1/2$ and $0$ for $x >0, =0$ and $<0$, respectively. 
Eq.(\ref{ctf}) is also called the local 
cutoff in light-front perturbative QCD\cite{Brodsky81}. All the 
$\Lambda$-dependences in the final bare Hamiltonian are removed by the 
counterterms. The use of eq.(\ref{ctf}) largely simplifies the analysis 
on the original cutoff scheme in ref.\cite{Wilson94}.

Meanwhile, for a practical SRG calculation, we also have to give an 
explicit form of the smearing function $f_{\lambda ij}$.  One of the 
simplest smearing functions that satisfies the requirements of SRG is 
a theta-function: \begin{equation}
	f_{\lambda ij} = \theta ({1\over 2} - x_{\lambda ij}). \label{sm1}
\end{equation}
On the light-front, it is convenient to redefine $x_{\lambda ij} =
{|P_i^- - P_j^-| \over P_i^- + P_j^- + \lambda^2/P^+}$ on the
light-front. Then we can 
further replace the above smearing function by the following form:
\begin{equation}
	f_{\lambda ij} = \theta({\lambda^2 \over P^+} - |\Delta P_{ij}^-| ), 
		\label{sm2}
\end{equation}
where $\Delta P_{ij}^- = P_i^- - P_j^-$ is the light-front free energy 
difference between the initial and final states of the physical 
processes.  The light-front free energies of the initial and final 
states are defined as sums over the light-front free energies of the 
constituents in the states. 

Note that the choice of a specific smearing function corresponds 
to picking up a particular renormalization scheme to specify the
renormalization scale in SRG. This is similar to the specification 
of the renormalization scale in usual RG by picking up a particular 
regularization scheme. As we will see later, physics must be 
independent of the choice of the smearing function. With the smearing 
function given by eq.(\ref{sm2}), eq.(\ref{eh11}) can be simply
reduced to
\begin{eqnarray}	
	H_{\lambda ij} &=& \theta({\lambda^2 / P^+} - |\Delta 
		P_{ij}^-| ) \Bigg\{ H_{ij}^B + \sum_k H^B_{Iik} 
		H^B_{Ikj} \Big[ \frac{\theta(|\Delta P_{ik}^-|-
		{\lambda^2/P^+}) \theta( |\Delta P_{ik}^-| - 
		|\Delta P_{jk}^-|)}{\Delta P^-_{ik}} \nonumber \\
		& & ~~~~~~~~~~~~~~~~~~~~~~~~~~~~~~~ 
		+ \frac{\theta(|\Delta P_{jk}^-|-
		{\lambda^2/P^+}) \theta( |\Delta P_{jk}^-| - 
		|\Delta P_{ik}^-|)}{\Delta P^-_{jk}} \Big]
		+ \cdots \Bigg\} .  \label{eh2}
\end{eqnarray}

\subsection{Application to light-front QCD}

Next, we shall calculate explicitly the low energy effective 
light-front QCD Hamiltonian in the similarity renormalization 
group scheme. Light-front QCD is the theory of QCD defined
on the light-front with the light-front gauge.  The canonical theory
of the light-front QCD was formulated in 70's \cite{Casher,Bardeen}.
In early 80's, Lepage and Brodsky developed the Hamiltonian formulation 
of light-front QCD and explored extensively the asymptotic behavior 
of hadrons and the perturbative expansion of various exclusive  
processes \cite{Brodsky81}. Here we shall use the two-component 
Hamiltonian formulation of the light-front QCD \cite{Zhang93b} that 
has been used in illustrating the ideas of the WCT scheme to 
nonperturbative QCD in the previous publication \cite{Wilson94}. 

The light-front coordinates are defined as follows: $x^\pm 
\equiv x^0 \pm x^3, x^i_\bot = x^i (i=1,2)$, with $x^+$ being chosen
as the light-front ``time'' direction along which the states
are evolved. Then $x^-$ and $x_\bot$ become naturally
the longitudinal and transverse coordinates. The inner product
of any two four-vectors is given by $a_\mu a^\mu = {1\over 2}
(a^+b^- + a^-b^+) - a_\bot \cdot b_\bot$, and the time and space 
derivatives ($\partial^\mu = {\partial \over \partial x_\mu}$)
and the 4-dimensional volume element are written as $\partial^+
= 2{\partial \over \partial x^-}$, $\partial^- = 2 {\partial \over 
\partial x^+}$, $\partial^i = -{\partial \over \partial x^i}$, 
and $d^4 x = {1\over 2} dx^+ dx^- d^2x_\bot$.

In the above coordinate system, the fermion field can be separated
by $\psi=\psi_+ + \psi_-$ with $\psi_\pm = {1\over 2} \gamma^0 
\gamma^\pm \psi$, and $\psi_+$ and $\psi_-$ are the so-called
dynamical and constraint components of the quark field on the 
light-front, respectively. Due to the gauge symmetry, only
two components of the 4-vector gauge field $A_a^\mu$ are dynamical 
field variables. With the choice of the light-front gauge 
$A_a^+=0$, we can explicitly solve the unphysical components of 
the quark and gauge fields ($A_a^-, \psi_-$) from the constraint 
equations which 
are obtained from QCD Lagrangian ${\cal L}_{QCD} = -{1\over 4} 
F^{\mu\nu}_a F_{a\mu\nu} + \overline{\psi} (i {\not \! \! D} 
+ m_q) \psi$:
\begin{eqnarray}
	&& \psi_- = \Big({1\over i\partial^+}\Big) (i\alpha_\bot 
		\cdot D_\bot + \beta m_q) \psi_+ , ~~~ \alpha_\bot = 
		\gamma^0 \gamma_\bot, ~~~ \beta = \gamma^0 \\
	&& A_a^- = 2\Bigg[\Big({1\over \partial^+}\Big) (\partial^i 
		A^i_a)+\Big({1\over \partial^+}\Big)^2(f^{abc}A_b^i
		\partial^+A_c^i + 2\psi_+ T_a \psi_+) \Bigg] , 
\end{eqnarray}
where  $T_a$ is the generator of SU(3) color group: $[T_a, T_b]
=if^{abc} T_c$.
Then the QCD Lagrangian on the light front can be expressed
in terms of only the physical field variables ($A_a^i, \psi_+$):
\begin{equation}
	{\cal L} = {1\over 2} (\partial^+ A_a^i \partial^- A_a^i)
		+ i \psi_+^\dagger \partial^- \psi_+ - {\cal H},
\end{equation}
where ${\cal H}$ is the light-front QCD Hamiltonian density
\cite{Zhang93b}:
\begin{equation}
	{\cal H} =  {\cal H}_0 + {\cal H}_I ,
\end{equation}
and
\begin{eqnarray}
	{\cal H}_0 &=& {1\over 2} (\partial^i A^j_a)(\partial^i A^j_a)
		+ \xi^{\dagger} {-\partial^2_{\bot} + m^2 \over 
		i\partial^+} \xi ,  \\
	{\cal H}_I &=& {\cal H}_{qqg} + {\cal H}_{ggg} + {\cal H}_{qqqq}
		+{\cal H}_{qqgg} + {\cal H}_{gggg} , \label{hdqcd}
\end{eqnarray}
with
\begin{eqnarray}
	& & {\cal H}_{qqg} = g \xi^{\dagger} \left\{ - 2 \left( 
		\frac{1}{\partial^+} \right) ( \partial \cdot 
		A_{\bot}) + \tilde{\sigma} \cdot A_{\bot} \left( 
		\frac{1}{\partial^+} \right) (\tilde{\sigma} 
		\cdot \partial_{\bot} + m) \right. \nonumber \\
	& & ~~~~~~~~~~~~~~~~~~~~~~ + \left. \left( \frac{1}{\partial^+} 
		\right) (\tilde{\sigma} \cdot \partial_{\bot} - m)
		\tilde{\sigma} \cdot A_{\bot} \right\} \xi , \label{qqg} \\
 	& & {\cal H}_{ggg} = g f^{abc} \left\{ \partial^i A_a^j A_b^i A_c^j
		+ (\partial^i A_a^i) \left( \frac{1}{\partial^+} \right)
		(A_b^j \partial^+ A_c^j) \right\} , \\
	& & {\cal H}_{qqgg} = g^2 \left\{ \xi^{\dagger} \tilde{\sigma} 
		\cdot A_{\bot} \left( \frac{1}{i \partial^+} \right) 
		\tilde{\sigma} \cdot A_{\bot} \xi \right. \nonumber \\
	& & ~~~~~~~~~~~~~~ \left. +~2  \left(\frac{1}{\partial^+} 
		\right)(f^{abc} A_b^i \partial^+ A_c^i) \left( \frac{1}{ 
		\partial^+} \right) (\xi^{\dagger} T^a \xi) \right\} , \\ 
	& & {\cal H}_{qqqq} = 2g^2 \left\{ \left(\frac{1}{\partial^+} 
		\right)(\xi^{\dagger} T^a \xi ) \left( \frac{1}
		{\partial^+} \right) (\xi^{\dagger} T^a \xi) \right\} , \\
	& & {\cal H}_{gggg} = \left. \frac{g^2}{4} f^{abc} f^{ade} 
		\right\{ A_b^i A_c^j A_d^i A_e^j \nonumber \\
	& & ~~~~~~~~~~~~~~~~~ \left. + 2 \left(\frac{1}{\partial^+} 
		\right)(A_b^i \partial^+ A_c^i) \left( \frac{1}{\partial^+} 
		\right) (A_d^j \partial^+ A_e^j) \right\}.  \label{gggg}
\end{eqnarray}
Here, we denoted $A_{\bot}=T^a A_{a\bot}, A_{a\bot}=(A_a^1, A_a^2)$.  
The notation $\tilde{\sigma}$ is defined by: $\tilde{\sigma}^1 = 
\sigma^2, \tilde{\sigma}^2 = - \sigma^1$ (the $2 \times 2$ Pauli 
matrices), and $\xi$ is
the two-component form of the light-front quark field:
\begin{equation}
	\psi_+ = \Lambda^+ \psi = \left[ \begin{array}{l} \xi \\ 
		0 \end{array} \right]~~,~~
	\psi_- =  \Lambda^- \psi =\left[ \begin{array}{c} 0 \\ 
		\left(\frac{1}{i\partial^+} \right)[\tilde{\sigma}^i
		(i \partial^i + gA^i) + im] \xi \end{array} \right] ,
\end{equation}
which comes from the use of the light-front $\gamma$-representation:
\begin{eqnarray}
	&& \gamma^0 = \left[\begin{array}{cc} 0 & -i \\ 
		i & 0 \end{array} \right]~~, ~~~~
	   \gamma^3 = \left[\begin{array}{cc} 0 & i \\ 
		i & 0 \end{array} \right] , \nonumber \\
	&& \gamma^1 = \left[\begin{array}{cc} -i \sigma^2 & 0 \\ 
		0 & i \sigma^2 \end{array} \right]~~, ~~~~
	   \gamma^2 = \left[\begin{array}{cc} i \sigma^1 & 0 \\ 
		0 & -i\sigma^1 \end{array} \right]. \label{lfgm}
\end{eqnarray}
Thus the bare light-front QCD Hamiltonian is given by
\begin{equation}
	H^B = \int_c dx^+ d^2 x_{\bot} \Big( {\cal H}_0 + 
		{\cal H}_I \Big)  + {\rm counterterms} , \label{qcdh}
\end{equation}
where $\int_c$ means that the local cutoff, eq.(\ref{ctf}), 
has been imposed. Also, counterterms are added to eq.(\ref{qcdh})
in order to remove all the $\Lambda$-dependence. Now, we can see 
that the canonical light-front QCD Hamiltonian is expressed purely 
in terms of the transverse gauge field components $A_\bot$ and 
the physical two-component quark fields $\xi$.

Upon the calculation to the second order in the coupling 
constant, the effective Hamiltonian (\ref{eh2}) in  
$q\overline{q}$ sector (with the initial and final states, 
$| i \rangle = b^{\dagger}(p_1, \lambda_1) 
d^{\dagger}(p_2,\lambda_2) | 0 \rangle$ and $| j \rangle = 
b^{\dagger}(p_3, \lambda_3) d^{\dagger}(p_4,\lambda_4) | 0 
\rangle$, respectively, where $p_i$ and $\lambda_i$ denote  
the respective momentum and  helicity of a quark on the 
light-front) becomes
\begin{eqnarray}
	H_{\lambda{ij}} &=& \theta({\lambda^2 / P^+} - |\Delta 
		P_{ij}^-| ) \Bigg\{ \langle j |:H^B:| i \rangle 
		- {g^2 \over 2\pi^2}\lambda^2 C_f {1 \over P^+}
		\ln \epsilon + {\rm mass ~ counterterms}  
		\nonumber \\
	& & ~~~~~~~~~~~~~ -g^2 (T^a)(T^a) {\theta(q^+) \over q^+} 
		\chi^{\dagger}_{\lambda_3} \left[2{ q^{i'}_{\bot}
		\over q^+} - {\tilde{\sigma} \cdot p_{3\bot}-im 
		\over [p_3^+]} \tilde{\sigma}^{i'} - \tilde{\sigma}^{i'} 
		{\tilde{\sigma} \cdot p_{1\bot} + im
		\over [p_1^+]} \right] \chi_{\lambda_1}\nonumber \\
	& & ~~~~~~~~~~~~~~~~~~~~~~\times \chi^{\dagger}_{-\lambda_2} 
		\Bigg[2{ q^{i'}_{\bot} \over q^+} - {\tilde{\sigma} 
		\cdot p_{2\bot} + im \over [p_2^+]} \tilde{\sigma}^{i'} 
		- \tilde{\sigma}^{i'} {\tilde{\sigma} \cdot 
		p_{4\bot} - im \over [p_4^+]} \Bigg] \chi_{-\lambda_4}
		F_{rij} \Bigg\} . \label{eh}
\end{eqnarray}
In eq.(\ref{eh}), $P^+= p_1^+ + p_2^+ = p_3^+ + p_4^+$, $\Delta P_{ij}^-
= p_1^- + p_2^- - p_3^- - p_4^-$, $:H^B:$ represents a normal ordering,
where the instantaneous interaction contribution to the quark 
self-energy has been included in the self-energy calculation which is
given by the mass counterterms and the logarithmic divergence, 
the color factor $C_f=(T^a T^a) = (N^2-1)/2N$, $N=3$ 
the total numbers of colors, and $\epsilon$ is an infrared longitudinal 
momentum cutoff.  Since $\ln \epsilon$ is an infrared 
divergence, it cannot be removed by mass counterterms. 
In  gauge symmetry, this divergence must be canceled in the 
physical sector (and this is true as we will see later). The last 
term in (\ref{eh}) corresponds to one transverse gluon exchange 
contribution in the second order of the low energy Hamiltonian 
expansion (\ref{eh2}).  The momentum $q$ is carried by 
the exchange gluon: $q^+ = p_1^+ - p_3^+ = p_4^+ - p_2^+$,
$q_{\bot} = p_{1\bot} - p_{3\bot} = p_{4\bot} - p_{2\bot}$, 
$\chi_{\lambda_i}$ denotes a helicity eigenstate.
The factor $F_{rij}$ arises from the similarity renormalization 
transformation:
\begin{eqnarray}
	F_{rij} &=& \Bigg\{\theta( {\Lambda^2/P^+} - |p_1^- - p_3^- 
		-q^-|) \theta( {\Lambda^2/P^+} - |p_4^- - p_2^- - q^-|)
		\nonumber \\
	& & ~~ \times \Bigg[{\theta(|p_1^- - p_3^- -q^-| 
		- {\lambda^2 \over P^+}) \theta( |p_1^- - p_3^- -q^-| - |p_4^- 
		- p_2^- - q^-|) \over p_1^- - p_3^- -q^- } \nonumber \\
	& &  ~~~~~~~ + {\theta(|p_4^- - p_2^- - q^-| - {\lambda^2 \over 
		P^+}) \theta(|p_4^- - p_2^- - q^-| - |p_1^- - p_3^- -q^-|)
		\over p_4^- - p_2^- - q^-} \Bigg] \nonumber \\
	& & ~~~~~~~ + (p_1 \longleftrightarrow p_3~, ~~ p_2 
		\longleftrightarrow p_4~, ~~ q \longrightarrow -q) 
		\Bigg\}.  \label{fcsr}
\end{eqnarray}
All light-front energies in eq.(\ref{fcsr}) are on mass-shell: 
$p_i^- = {p_{i\bot}^2 + m_i^2 \over p_i^+}$ and $q^- =
{q_{\bot}^2 \over q^+}$.   

If we continue to evaluate all the terms in the expansion of
eq.(\ref{eh2}), the resulting Hamiltonian is the exact 
QCD Hamiltonian at the energy scale $\lambda$. In practice, 
we only consider the 
leading and the next-to-leading terms, i.e., eq.(\ref{eh}), as a
starting effective Hamiltonian. The basic idea to realize
a weak-coupling treatment of QCD for hadrons is whether we can 
solve hadron states from this effective Hamiltonian (\ref{eh}) 
with an arbitrary small coupling constant $g$ such
that the higher order corrections in (\ref{eh2}) can be handled 
perturbatively. From the success of the constituent quark model,
we understand that a necessity for such a realization might 
depend on the existence of a confining interaction in (\ref{eh}).

\subsection{Quark confining interaction on the light-front}

Naively, we know that any weak-coupling Hamiltonian 
derived from QCD will have only Coulomb-like interactions, and 
confinement can only be exhibited in a strong-coupling theory.
However, Perry has recently found that upon to the 
second order calculation of the low energy Hamiltonian, 
a logarithmic confining potential has already occurred 
\cite{Perry94}.  Explicitly, the two-body 
quark-antiquark interaction from the first term of 
eq.(\ref{eh}) is the instantaneous gluon exchange 
interaction ${\cal H}_{qqqq}$ which has the form in
the momentum space: 
\begin{equation} 
	- {1 \over (q^+)^2} .
\end{equation}
The confinement must be associated with the interaction where 
$q^+ \rightarrow 0$.  This is because only these particles with 
zero longitudinal momentum can occupy the 
light-front vacuum state\cite{Wilson94}.  Thus, we need 
to analyze the feature of the effective Hamiltonian 
when $q^+ \rightarrow 0$.  For $q^+ \rightarrow 0$, the dominant  
second order contribution from one transverse gluon 
exchange interaction in eq.(\ref{eh}) is given by
\begin{equation}	\label{ogx}
    \Bigg[{1 \over (q^+)^2} {q^2_{\bot} \over q^2_{\bot} } + 
	O({1 \over q^+})\Bigg] \theta (q^- - \lambda^2/P^+) . 
\end{equation}
In the usual perturbative calculation, no such a $\theta$-function 
(coming from the smearing function) is attached in the above equation.
Thus these two dominant 
contributions from the instantaneous and one-gluon exchange 
interactions are exactly cancelled when $q^+ \rightarrow 0$.  Only 
Coulomb-type interaction (the terms $\sim O({1\over q^+})$) remains. 
However, in the light-front similarity renormalization group scheme, 
the corresponding one-gluon exchange contribution of eq.(\ref{ogx}) 
only contains these gluons
with energy being greater than the energy cutoff $\lambda^2 /P^+$.
As a result, the instantaneous gluon exchange term $1/(q^+)^2$ 
remains uncancelled if the gluon energy, $q^2_{\bot} /q^+$, is 
less than $\lambda^2 /P^+$.  The remaining uncancelled instantaneous 
interaction contains an infrared divergence and a finite part 
contribution (for a detailed derivation, see the next section). 
The divergence part is cancelled precisely for physical states 
by the same divergence in the quark self-energy correction
[see (\ref{eh})]. The remaining
finite part corresponds to a logarithmic confining potential:
\[  b_{\lambda} \ln (P^+|x^-|) + c_{\lambda} \ln(\lambda^2 
	|x_{\bot}|^2) . \]
The above result is indeed first obtained by Perry with the 
use of the concept 
of coupling coherence and a slightly different renormalization 
scheme \cite{Perry94} (also see a oversimple derivation given
by Wilson \cite{Wilson94a}).  Here the derivation is purely 
based on the light-front similarity renormalization scheme.  

One may argue that the existence of such a confining potential 
in $H_{\lambda}$ up to the second order may only be an artificial effect 
designed in the renormalization scheme we used here.  If we included
the interaction with the exchange gluon energy below the cutoff,
then the instantaneous interaction would be completely cancelled,
and no such confining potential should exist, as expected in
the usual perturbation computation. Wilson has pointed out that the 
set up of the new renormalization scheme is motivated by the idea
that the gluon mass must be nonzero in the low energy domain
(a constituent picture). The massive gluon can be originated 
from the nonlinear interactions in non-abelian gauge theory. 
Therefore, the non-cancellation of the instantaneous interaction 
in the low energy domain is indeed a consequence of the existence 
of the constituent massive gluons due to the non-abelian gauge 
interactions.  It is independent of any particular renormalization scheme.
The use of the low energy cutoff $\lambda$ (namely, taking the
smearing function as a theta function) just gives us a simple 
realization of this confining picture that the exchange energy
of massive dynamical gluons cannot run down to the zero value in 
nonperturbative QCD. 

Yet, although there exists a confining interaction in $H_\lambda$
even up to the second order, this effective QCD Hamiltonian is 
already very complicated, due to the spin dependent part in the
interactions. As we know the spin dependent interactions on the 
light-front are essentially related to the chiral symmetry breaking. 
In order to examine the above picture of confining mechanism on
the light-front and to develop explicitly a weak-coupling treatment 
approach of nonperturbative QCD to hadronic bound states, in the 
next section we shall utilize the above formulation to heavy
quark systems. We find that the low-energy QCD Hamiltonian  
$H_{\lambda}$ for heavy quarks can be largely simplified and 
an analytic form consisting of the confining  
and Coulomb potentials emerges.

\section{Heavy quark confining Hamiltonian}
 
In the past few years, QCD has been made a numerous progresses in
understanding the heavy hadron structure, due mainly to the discovery 
of heavy quark symmetry by Isgur and Wise \cite{Isgur90}, and the 
development of HQET by Georgi et al. \cite{Georgi}.  However, to 
completely understand the nonperturbative QCD dynamics of
heavy hadrons, one should solve heavy hadron bound states directly 
from QCD, which is still an open question. Next we will use 
SRG to the light-front HQET to derive a heavy quark 
confining Hamiltonian, from which we may solve from QCD the heavy 
hadron bound states directly.

\subsection{Light-front HQET}

The light-front heavy quark effective Lagrangian derived from
QCD lagrangian ${\cal L} = \overline{Q}(i {\not \! \! D}
-m_Q) Q$ as a $1/m_Q$ expansion is given in 
refs.\cite{Zhang95,Cheung95}: 
\begin{eqnarray}
       && {\cal L} = {2 \over v^+} {\cal Q}_{v+}^{\dagger} (iv \cdot D)
                {\cal Q}_{v+} - \sum_{n=1}^{\infty} \Big({ 1 \over m_Q v^+}
                \Big)^n {\cal Q}_{v+}^{\dagger} \Big\{(i\vec{\alpha}
                \cdot \vec{D})(-i D^+)^{n-1}(i\vec{\alpha} \cdot 
		\vec{D}) \Big\} {\cal Q}_{v+}, 
\end{eqnarray}
where ${\cal Q}_{v+}$ is the light-front dynamical component 
of the heavy quark field after the phase redefinition: $Q(x)= 
e^{-im_Q v \cdot x} ({\cal Q}_{v+}(x) + {\cal Q}_{v-}(x))$,
$v^{\mu}$ the four velocity of the heavy hadrons, $P^\mu
=M_H v^\mu$ with $v^2=1$ and $M_H$ the heavy hadron
mass, $ \vec{\alpha} \cdot \vec{D} \equiv \alpha_{\bot} 
\cdot D_{\bot} - {1\over v^+} (\alpha_{\bot} \cdot v_{\bot} + 
\beta) D^+$, and $D^{\mu}$ the usual covariant derivative.  
The corresponding light-front heavy quark bare Hamiltonian 
density is given by
\begin{eqnarray}
    && {\cal H} = { 1\over iv^+} {\cal Q}^{\dagger}_{v+} (v^-\partial^+
                -2v_{\bot} \cdot \partial_{\bot} ) {\cal Q}_{v+}
                - {g \over v^+} {\cal Q}^{\dagger}_{v+} (v \cdot A)
                {\cal Q}_{v+} \nonumber \\
	 & & ~~~~~~~~~~~ + \sum_{n=1}^{\infty} \Big({ 1 \over m_Q v^+}
                \Big)^n {\cal Q}_{v+}^{\dagger} \Big\{(i\vec{\alpha}
                \cdot \vec{D}) (-i D^+)^{n-1} (i \vec{\alpha} \cdot 
		\vec{D}) \Big\} {\cal Q}_{v+}.  \label{hqbeh}
\end{eqnarray}
The heavy antiquark Hamiltonian has the same form except for 
replacing $v$ by $-v$.

In the large $m_Q$ limit, only the leading (spin and mass 
independent) Hamiltonian is remained. The $1/m_Q^n$ terms 
($n \geq 1$) in (\ref{hqbeh}) can 
be regarded as perturbative corrections to the leading order 
operators and states. To determine 
confining interactions in heavy quark systems, the leading 
heavy quark Hamiltonian plays an essential role.  
With the light-front gauge $A^+=0$, the leading-order 
bare QCD Hamiltonian density is 
\begin{eqnarray}
    {\cal H}_{ld} &=& { 1\over iv^+} {\cal Q}^{\dagger}_{v+} (v^-\partial^+
                -2v_{\bot} \cdot \partial_{\bot} ) {\cal Q}_{v+} 
     		- {2g \over v^+} {\cal Q}^{\dagger}_{v+} \left\{v^+ \Big[
		\Big({1\over \partial^+}\Big) \partial_{\bot} \cdot A_{\bot}
		\Big] - v_{\bot} \cdot A_{\bot} \right\} {\cal Q}_{v^+}
		\nonumber \\
     & & ~~~~~~ + 2g^2 \Big({1\over \partial^+}\Big) \Big({\cal 
		Q}^{\dagger}_{v+} T^a {\cal Q}_{v^+} \Big) \Big({1 \over
		\partial^+} \Big) \Big( \psi_+^\dagger T^a 
		\psi_+ \Big) , \label{ldhh}
\end{eqnarray}
where $\psi_+$ is either the heavy antiquark field or the 
light-front quark field operator in the present consideration. 
Note that besides the leading term in eq.(\ref{hqbeh}), 
the above bare Hamiltonian has also already included 
the relevant terms from the gauge field part, $-{1\over 2}
{\rm Tr} (F_{\mu\nu} F^{\mu\nu})$, of the QCD Lagrangian.  
These terms come from the elimination of the unphysical gauge 
degrees of freedom, the longitudinal component $A^-_a$\cite{Zhang93b}.
Eq.(\ref{ldhh}) has obviously the spin and flavour heavy
quark symmetry, or simply the heavy quark symmetry.

The above leading Hamiltonian (or Lagrangian) is the basis 
of the QCD-based description for heavy hadrons containing a single 
heavy quark, such as $B$ and $D$ mesons.  As recently pointed 
out by Mannel et al.\cite{Mannel93,Mannel95} the purely heavy
quark leading Lagrangian may be not appropriate to describe
heavy quarkonia.  This is because the 
anomalous dimension of QCD radiative correction to 
$Q\overline{Q}$ currents contains an infrared singularity 
in the limit of two heavy constituents having equal
velocity. Such an infrared singularity is a long distance 
effect and should be absorbed into quarkonium states.  
To solve this problem, they show that one should incorporate 
the effective Hamiltonian with the first order kinetic 
energy term into the leading Hamiltonian \cite{Mannel95}. The
light-front kinetic energy can be obtained from eq.(\ref{hqbeh}),
\begin{equation}
	{\cal H}_{kin} = - {1\over m_Q v^+} {\cal Q}_{v+}^\dagger
		\Bigg\{ \partial_\bot^2 - {2 v_\bot \cdot 
		\partial_\bot \over v^+}
		\partial^+ + {v^- \over v^+} \partial^{+2} \Bigg\}
		{\cal Q}_{v+}.	\label{nloh}
\end{equation}
Indeed, as we will see in Section IV.C, for heavy quarkonia 
the heavy quark carries a relative momentum that is proportional 
to heavy quark mass. Then the first order kinetic energy term is 
of the same order as the leading Hamiltonian. Other terms in the 
$1/m_Q$ are really suppressed by the power of $1/m_Q$. This is why 
one have to incorporate the leading Hamiltonian with the first 
order kinetic energy term. As a consequence, in the heavy mass 
limit, quarkonia have spin symmetry but no flavour symmetry.

\subsection{Confining Hamiltonian for heavy quarkonia}

Within light-front HQET, we now follow the procedure 
described in the previous section to find an effective QCD 
Hamiltonian for $Q\overline{Q}$ systems. The bare Hamiltonian 
for $Q\overline{Q}$ systems contains (\ref{ldhh}) and (\ref{nloh}) 
for both heavy quark and antiquark plus the full QCD Hamiltonian
for gluons and light quarks\cite{Zhang93b}. Since the kinetic energy 
(\ref{nloh}) is of order $\Lambda_{QCD}/m_Q$, which is at most of
the same order of the Coulomb interaction, we can treat the kinetic 
energy in the same way as the instantaneous $Q\overline{Q}$ interaction 
[the last term in eq.(\ref{ldhh})]. Thus, the free Hamiltonian 
$H_0$ used in SRG is given only by the first term in eq.(\ref{ldhh}) 
plus the free gluon Hamiltonian.

Keeping the above consideration in mind, it is easy to compute the
effective Hamiltonian eq.(\ref{eh2}) for $Q\overline{Q}$ systems.
Following the WCT ideas, we shall  
calculate $H_\lambda$ for $Q\overline{Q}$ systems up to the 
second order in the initial and final states 
defined by $| i \rangle = b^{\dagger}_v(k_1, \lambda_1) 
d^{\dagger}_v(k_2,\lambda_2) | 0 \rangle$ and $| j \rangle = 
b^{\dagger}_v(k_3, \lambda_3) d^{\dagger}_v(k_4,\lambda_4) 
| 0 \rangle$, respectively, where $k_i$ is the residual momentum 
of heavy quarks, $p^{\mu}_i= m_Q v^{\mu} + k^{\mu}_i$, 
and $\lambda_i$ its helicity.  The operator $b^{\dagger}_v(k_i,
\lambda_i) ~[d^{\dagger}_{-v}(k_i, \lambda_i)]$ creates a heavy quark 
[antiquark] with velocity $v$, 
\begin{equation}
	\{b_v(k,\lambda), ~ b^{\dagger}_{v'}(k',\lambda')\} 
		=\{d_{-v}(k,\lambda), ~ d^{\dagger}_{-v'}(k',\lambda')\}
		= 2 (2\pi)^3 \delta_{vv'} \delta_{\lambda \lambda'}
		\delta^3 (\bar{k}-\bar{k}'),
\end{equation}
and $\delta^3(\bar{k}-\bar{k}') \equiv \delta(k^+-{k'}^+) 
\delta^2(k_\bot-k'_\bot)$.  The result is 
\begin{equation}
	H_{\lambda 0 ij} = H_{Q\overline{Q}free ij}  
		+ V_{Q\overline{Q} ij}  , \label{QQeh}
\end{equation}
where
\begin{eqnarray}
	H_{Q\overline{Q}free ij} &&= [2(2\pi)^3]^2 \delta^3
		(\bar{k}_1-\bar{k}_3) \delta^3(\bar{k}_2-\bar{k}_4) 
	 	\delta_{\lambda_1 \lambda_3} \delta_{\lambda_2 
		\lambda_4} \nonumber \\
	& & ~~~~~~~~~~~~~~~ \times \Bigg\{ {\overline{\Lambda}\over m_Q}
	 	\Big[2\kappa_\bot^2 + \overline{\Lambda}^2(2y^2-2y +1)
		\Big] - \overline{\Lambda}^2 - 2{g^2 \over 4\pi^2} 
		C_f {\lambda^2 \over K^+}\ln \epsilon \Bigg\} , 
			\label{QQfe} \\
	V_{Q\overline{Q} ij} (y-y', && \kappa_\bot -\kappa'_\bot) 
	     =  2(2\pi)^3 \delta^3(\bar{k}_1 + \bar{k}_2 - \bar{k}_3 - 
		\bar{k}_4) \delta_{\lambda_1 \lambda_3} \delta_{\lambda_2 
		\lambda_4}  \nonumber \\
	&& ~~~~~ \times {-4g^2 (T^a)(T^a) \over (K^+)^2} \Bigg\{{1 
		\over (y-y')^2} \Big(1 - \theta( A(y-y', \kappa_\bot 
		- \kappa'_\bot, \overline{\Lambda}) - \lambda^2) \Big)
		\nonumber \\ &&
	 ~~~~~~~~~~~~ + {\overline{\Lambda}^2 \over (\kappa_{\bot} - 
		\kappa_{\bot}')^2 + (y-y')^2 \overline{\Lambda}^2} 
		\theta(A(y-y', \kappa_\bot - \kappa'_\bot, 
		\overline{\Lambda}) - \lambda^2) 
		\Bigg\}  \label{QQvv}.
\end{eqnarray}
Here we have introduced the longitudinal residual momentum 
fractions and the relative transverse residual momenta, 
\begin{eqnarray}
	&& y = k^+_1 / K^+ ~,~~~~~ \kappa_{\bot} = 
		k_{1\bot} - y K_{\bot} , \nonumber \\
	&& y' = k^+_3 / K^+ ~,~~~~~ \kappa'_{\bot} = 
		k_{3\bot} - y' K_{\bot},	\label{QQlmf}
\end{eqnarray}
where $K^\mu$ is defined as the residual center mass momentum 
of the heavy quarkonia: $K^\mu = \overline{\Lambda} v^{\mu}$,
and $\overline{\Lambda} = M_H - m_{Q} - m_{\overline{Q}}$ is a 
residual heavy hadron mass. It follows that $K^+ = k_1^+ + k_2^+ 
= k_3^+ + k_4^+$, $K_{\bot}= k_{1\bot} + k_{2\bot} = k_{3\bot} 
+ k_{4\bot}$. Since  $ 0 \leq p_1^+ = m_Q v^+ + k_1^+ \leq M_Hv^+$,
in the heavy quark mass limit, we have $M_H \rightarrow 2m_Q$ 
so that $-m_Qv^+ \leq k^+_1, ~ k^+_3 \leq m_Q v^+$.  Hence, 
the range of $y$ and $y'$ are given by $- \infty < y , y' < \infty $.
We have also defined in eq.(\ref{QQvv})
\begin{equation}
	A(y-y', \kappa_\bot - \kappa'_\bot, \overline{\Lambda}) 
		\equiv {(\kappa_\bot - \kappa'_\bot)^2 \over |y-y'|} 
		+ |y-y'| \overline{\Lambda}^2 .
\end{equation}

Eq.(\ref{QQeh}) is the nonperturbative part of the effective 
Hamiltonian for heavy quarkonia in the WCT scheme, in which 
we have already let UV cutoff parameter $\Lambda \rightarrow 
\infty$ and the associated divergence has been put in the mass 
correction. The kinetic energy (\ref{nloh}) is now included 
in the above effective Hamiltonian [the $1/m_Q$ term in 
eq.(\ref{QQfe})]. Note that there is an infrared divergent 
term in eq.(\ref{QQfe}) which comes from the quark self-energy 
correction in SRG,
\begin{equation}
	2\Sigma  = -{g^2 \over 2\pi^2}\lambda^2 C_f \ln \epsilon 
		+ 2 \delta m_Q^2 ,	\label{cqse}
\end{equation}
where $\epsilon$ is an infrared cutoff of the momentum fraction
$q^+/K^+$, and $q^+$ the longitudinal momentum carried by gluon
in the quark self-energy loop. The usual mass correction 
$\delta m_Q^2 = {g^2 \over 4\pi^2} C_f \overline{\Lambda}^2\ln 
{\Lambda^2 \over \lambda^2}$, has been renormalized away in 
eq.(\ref{QQfe}), where the color factor $C_f=(T^a T^a) = (N^2-1)/2N$, 
$N=3$ the total numbers of colors. In the WCT scheme, by removing 
away this mass correction, we should assign the corresponding
constituent quark mass in $H_{\lambda 0}$ being $\lambda$-dependent.  
But, the heavy quark mass is larger than the low energy scale.  
Its dependence on $\lambda$ should be very weak and could be 
neglected. Meanwhile, the $Q\overline{Q}$ interaction (\ref{QQvv})
contains two contributions: the instantaneous interaction plus 
the second order contribution in eq.(\ref{eh2}) [i.e.  
the terms proportional to the theta function in eq.(\ref{QQvv})].
We shall show next that the above $V_{Q\overline{Q}}$ is indeed a 
combination of a confining interaction plus a Coulomb-type 
interaction.

\subsection{A simple picture of quark confinement on the light-front}

In the conventional picture, QCD has a complex vacuum that contains 
infinite quark pairs and gluons necessary for confinement and chiral 
symmetry breaking.  On the light-front, the longitudinal momentum 
of physical particles is always positive, $p^+ = p^0 + p^3 \geq 0$.
Consequently, only these constituents with zero longitudinal momentum 
(called zero modes \cite{Burkardt}) can occupy the light-front vacuum.  
The zero modes carry an extremely high light-front energy which 
has been integrated out in SRG.  As a result, some equivalent
effective interactions are generated in $H_\lambda$ so that 
the light-front QCD vacuum becomes trivial.  The nature 
of nontrivial QCD vacuum structure, the confinement as well as 
the chiral symmetry breaking, is then made manifestly in 
$H_\lambda$ in terms of these new 
effective interactions. We will see that $H_{\lambda 0}$
explicitly contains a confining interaction at long distances.
The interactions associated with the chiral 
symmetry breaking may be manifested in the fourth order 
computation of $H_\lambda$ for light quark systems
\cite{Wilson94a}, but these interactions are not important 
in the study of heavy hadrons here. 

The confining interaction can be easily obtained by applying the
Fourier transformation to the first term in (\ref{QQvv}).  
It is convenient to perform the calculation in the frame
$K_\bot =0$, in which
\begin{eqnarray}
	 \int {dq^+ d^2q_\bot \over (2\pi)^2} && e^{i(q^+ x^- 
		+ q_\bot \cdot x_\bot)} \Bigg\{-{4g_\lambda^2 
		\over K^{+2}}~{1 \over (y-y')^2} \theta(\lambda^2 
		- A(y-y', \kappa_\bot - \kappa'_\bot, 
		\overline{\Lambda})) \Bigg\}  \nonumber \\
	&&= - {g_\lambda^2 \over 2\pi^2} \int_0^{{\lambda^2 \over 	
		\overline{\Lambda}^2} K^+} dq^+ e^{iq^+ x^-}{q^2_{\bot m} 
		\over q^{+2}}~{2J_1(|x_\bot|q_{\bot m}) \over 
		|x_\bot| q_{\bot m}},		\label{conp}
\end{eqnarray}
where $q^+ = k_1^+-k_3^+=K^+(y-y'), q_\bot = k_{1\bot} - k_{3\bot} =
\kappa_\bot - \kappa'_\bot$ for $K_\bot =0$, $q_{\bot m} \equiv 
\sqrt{{\lambda^2 \over K^+} q^+ - {\overline{\Lambda}^2 \over 
K^{+2}} q^{+2}}$, and $J_1 (x)$ is a Bessel function.   
An analytic solution to the
integral (\ref{conp}) may be difficult to carry out. 
However, the nature of confining interactions is a large 
distance QCD behavior.  We may consider the integral
for large $x^-$ and $x_\bot$.  In this case, if $q^+ x^-$ and/or 
$|x_\bot|q_{\bot m}$ are large, the integration vanishes,
yet $J_1(x) = {x\over 2} + {x^3\over 16} + \cdots$ for
small $x$.  The dominant contribution of the integral 
(\ref{conp}) for large $x^-$ and $x_\bot$ comes from the small 
$q^+$ such that $q^+ x^-$ and/or $|x_\bot|q_{\bot m}$ must 
remain small, which leads to $e^{iq^+ x^-}{2J_1(|x_\bot|q_{\bot m}) 
\over |x_\bot| q_{\bot m}} \simeq 1 $. This corresponds to 
$q^+ < {1 \over x^-}$ and/or $q^+ < {K^+ \over |x_\bot|^2 
\lambda^2 }$.

If $q^+ < {1 \over x^-}< { K^+ \over |x_\bot|^2 \lambda^2 }$, 
eq.(\ref{conp}) is reduced to
\begin{equation}
	-{g_\lambda^2 \over 2\pi^2} \int_0^{1\over x^-} dq^+  
		{1\over q^{+2}}\Bigg({\lambda^2 \over K^+} q^+ - 
		{\overline{\Lambda}^2 \over K^{+2}} q^{+2} \Bigg) 
	    = {g_\lambda^2\lambda^2\over 2\pi^2 K^+} 
		 \Big(\ln (K^+|x^-|) + \ln \epsilon \Big) , \label{conp1}
\end{equation}
where a term $\sim {1 \over x^-}$ is neglected since $x^-$ is large,
and $\epsilon$ is an infrared cutoff of the momentum fraction 
$q^+/K^+$. It is the same as in eq.(\ref{cqse}) so that the above
 infrared logarithmic divergence ($\sim \ln \epsilon$) exactly 
cancels the divergence in eq.(\ref{QQfe}) for color single states.
What remains is a logarithmic confining interaction 
(except for a color factor):
\begin{equation}
	V_{conf.}(x^-, x_\bot) \sim {g_\lambda^2 \lambda^2\over 
		2\pi^2 K^+} \ln (K^+|x^-|).  \label{lconp}
\end{equation}
Similarly, when  $q^+ < { K^+ \over |x_\bot|^2 \lambda^2 }< 
{1 \over x^-}$, we have
\begin{equation}
	-{g_\lambda^2 \over 2\pi^2} \int_0^{K^+ \over 
		|x_\bot|^2 \lambda^2} dq^+ {1\over q^{+2}}\Big({\lambda^2
		\over K^+} q^+ - {\overline{\Lambda}^2 \over 
		K^{+2}} q^{+2} \Big) = {g_\lambda^2 \lambda^2\over 
		2\pi^2 K^+} \Big(\ln ( \lambda^2 |x_\bot|^2 ) + 
		\ln \epsilon \Big) ,	\label{conp2}
\end{equation}
where the term $\sim {1 \over x^2_\bot}$ has also been ignored 
because of the large $x^2_\bot$. Again, the infrared divergence 
($\sim \ln \epsilon$) is cancelled in $H_\lambda$ for physical
states, and we obtain the following confining interaction:
\begin{equation}
	V_{conf.}(x^-, x_\bot) \sim {g_\lambda^2 \lambda^2\over 
	 2\pi^2 K^+} \ln (\lambda^2 |x_\bot|^2 ).  \label{tconp}
\end{equation}
Hence, the effective Hamiltonian $H_{\lambda 0}$ exhibits a logarithmic 
confining interaction between a heavy quark and a heavy antiquark 
in all the directions of $x^-$ and $x_\bot$ space.  

The Coulomb interaction corresponds the second term in (\ref{QQvv}),
its Fourier transformation (except for the color factor) is
\begin{equation}
	{\overline{\Lambda}^2  \over (\kappa_{\bot} - 
		\kappa_{\bot}')^2 + (y-y')^2 \overline{\Lambda}^2} 
		\sim  {1 \over 4\pi} \int dx^- d^2x_\bot e^{i(x^-q^+
		+ q_\bot \cdot x_\bot)}
		\Bigg({\overline{\Lambda} \over K^+} \Bigg) {1 \over 
		r_l} ,	\label{Coulp}
\end{equation}
where $r_l \equiv 
\sqrt{x^2_\bot + \Big({\overline{\Lambda} \over K^+} \Big)^2 
(x^-)^2}$ which is defined as a ``radial'' variable in the light-front 
space\cite{Wilson94}. Eq.(\ref{Coulp}) shows that the Coulomb 
interaction on the light-front has the form
\begin{equation}
	V_{Coul.}(x^-, x_\bot) \sim  - {g_\lambda^2 \over 4\pi} 
		{\overline{\Lambda} \over K^+} {1 \over r_l}.
\end{equation}
Thus, we have explicitly shown that $H_{\lambda 0}$ contains a 
Coulomb interaction at short distances and a confining interaction 
at long distances.
 
Moreover, a clear light-front picture of quark confinement 
emerges here. To be specific, we define quark confinement 
as follows: i) There is a 
confining interaction between quarks such that quarks cannot be 
well-separated; ii) No color non-singlet bound states exist 
in nature, only color singlet states with finite masses can 
be produced and observed; and iii) The conclusions of i--ii)
are only true for QCD but not for QED.  

We have shown explicitly the existence of a confining 
interaction in $H_{\lambda 0}$. One can also easily see 
from $H_{\lambda 0}$ the non-existence of color non-singlet 
bound states.  This is essentially related to the infrared 
divergences in $H_{\lambda 0}$. From eqs.(\ref{conp1}) and 
(\ref{conp2}), we find that the uncancelled instantaneous 
interaction contains a logarithmic infrared divergence.
Except for the color factor, this infrared divergence has the 
same form as the divergence in eq.(\ref{QQfe}).
Thus, we immediately obtain the following conclusions.

	(a). For a single (constituent) quark state, the 
interaction part of $H_{\lambda 0}$ does not contribute to
its energy. The remaining infrared divergence from quark
self-energy correction implies that the dynamical quark
mass for a single quark state is infinite (infrared divergent) 
and cannot be renormalized 
away in the spirit of gauge invariance.  Equivalently speaking,
single quark states carry an infinitely large mass and therefore
they cannot be produced. 

	(b). For color non-singlet composite states, the color 
factor $(T^a)_{\alpha \beta} (T^a)_{\delta \gamma}$ in the 
$Q\overline{Q}$ interaction is different from the color factor 
$C_f=\tr(T^a T^a)$.
Therefore, the infrared divergence in the self-energy correction
also cannot be cancelled by the corresponding divergence
from the uncancelled instantaneous interaction. As a result,
color non-singlet composite states are infinitely heavy
that they cannot be produced as well.

	(c). For color singlet $Q\overline{Q}$ states, the 
color factor $(T^a)(T^a) \rightarrow C_f$.  Thus, 
the infrared divergences are completely cancelled and the resulting
effective Hamiltonian is finite.  In other words, only color 
singlet composite are physically observable.

Finally, we argue that the above mechanism of quark 
confinement is indeed only true for QCD.  As we have
seen the light-front confinement interaction is just an 
effect of the non-cancellation between instantaneous 
interaction and one transverse gluon interaction generated
in SRG. Such a non-cancellation arises in SRG because we 
introduce the energy scale $\lambda$ through the smearing 
function. Introducing the energy
scale $\lambda$ in SRG forces the transverse gluon energy 
involved in the $Q\overline{Q}$ effective interaction 
never be less than a certain value (the energy scale $\lambda$). 
This implies that the gluon may become
massive at the hadronic mass scale.  Of course, such a gluon mass 
must be a dynamical mass generated from the highly nonlinear gluon 
interactions. In other words, the above confining picture is 
indeed a dynamical consequence of non-Abelian gauge theory. This 
confinement mechanism is not valid in QED.  In QED, since  
photon mass is always zero, the photon energy covers the entire 
range from zero to
infinity.  Thus, in QED, we can always choose the energy 
scale $\lambda$ being zero. With $\lambda=0$, the infrared 
divergences do not occur in the electron self-energy 
correction.  As a result, the renormalized single electron mass
is finite, in contrast to the divergent mass of single quark 
states.  For the same reason, with $\lambda=0$, 
the instantaneous interaction in
the effective QED Hamiltonian is also exactly cancelled
by the same interaction from one transverse photon exchange  
so that only one photon exchange Coulomb interaction remains.  
Thus, applying SRG to QED  and let $\lambda =0$ in 
the end of procedure, we obtain a conventional QED 
Hamiltonian which only contains the Coulomb interaction.  

\subsection{Generality of the confining Hamiltonian in SRG}

As we have pointed out in the previous section, the existence of
quark confining interaction in this formalism of nonperturbative
QCD should be independent of the choice of a particular 
renormalization scale. In this subsection, 
we shall examine if such a light-front confining picture 
could be an artificial effect designed in our SRG renormalization 
scheme, especially if it is obtained from our specific 
choice of the smearing function (\ref{sm2}).  To answer this 
question, we shall rederive $H_\lambda$ without specifying
the detailed form of the smearing function $f_{\lambda ij}$. 

For heavy quarkonia in the HQET, $\Delta K^-_{ik} = \Delta K^-_{jk}
= -{1\over K^+(y-y')} \Big((\kappa_\bot-\kappa'_\bot)^2 +
(y-y')^2\overline{\Lambda}^2 \Big)$. This
immediately leads to $f_{\lambda ik} = f_{\lambda jk}$ because 
$x_{\lambda ik} = x_{\lambda jk} = {|\Delta K^-_{ik}| \over K_i^-
+ K^-_j - \Delta K^-_{ik} + \lambda^2/K^+}$. Thus eq.(\ref{sd1})
is reduced to 
\begin{equation}
	g_{\lambda ijk} = 1 - f_{\lambda ik}^2 = 1 - f_{\lambda jk}^2.
\end{equation}
Following the same calculation of Sec.III.B, we can find that
\begin{equation}
	H_{\lambda 0 ij} = f_{\lambda ij} \Big(H_{Q\overline{Q}free ij}  
		+ V_{Q\overline{Q} ij} \Big) , \label{QQeh1}
\end{equation}
where
\begin{eqnarray}
	&& H_{Q\overline{Q}free ij}= [2(2\pi)^3]^2 \delta^3
		(\bar{k}_1-\bar{k}_3) \delta^3(\bar{k}_2-\bar{k}_4) 
	 	\delta_{\lambda_1 \lambda_3} \delta_{\lambda_2 
		\lambda_4} \nonumber \\
	& & ~~~~~~~~~~~~~~~~~~~~~~~~~~~~ 
		\times \Bigg\{ {\overline{\Lambda}\over m_Q}
	 	\Big[2\kappa_\bot^2 + \overline{\Lambda}^2(2y^2-2y +1)
		\Big] - \overline{\Lambda}^2 - 2{g^2 \over 4\pi^2} 
		C_f {\lambda^2 \over K^+} f_s \ln \epsilon \Bigg\} , 
			\label{QQfe1} \\
	&& V_{Q\overline{Q} ij} (y-y', \kappa_\bot -\kappa'_\bot) 
	     =  2(2\pi)^3 \delta^3(\bar{k}_1 + \bar{k}_2 - \bar{k}_3 - 
		\bar{k}_4) \delta_{\lambda_1 \lambda_3} \delta_{\lambda_2 
		\lambda_4}  \nonumber \\
	&& ~~~~~~~~ \times {-4g^2 (T^a)(T^a) \over (K^+)^2} \Bigg\{{1 
		\over (y-y')^2} ~ f_{\lambda ik}^2 
		 + {\overline{\Lambda}^2 \over (\kappa_{\bot} - 
		\kappa_{\bot}')^2 + (y-y')^2 \overline{\Lambda}^2} 
		(1 -  f_{\lambda ik}^2 ) \Bigg\}  \label{QQvv1}.
\end{eqnarray}
In Eq.(\ref{QQfe1}), $f_s$ depends on the detailed form of $f_{\lambda ik}$.

If we took the smearing function $f_{\lambda ik}$ by the theta function
of Eq.(\ref{sm2}), then Eq.(\ref{QQvv1}) will immediately be reduced to
Eq.(\ref{QQvv}). Now we shall consider the general case. Note that
$K^-_i + K^-_j = 2\overline{\Lambda} v^- << \lambda^2/K^+$ since 
$\overline{\Lambda}$ is of the same order $\Lambda_{QCD} \simeq 0.1
\sim 0.4$ GeV (for heavy quarkonia, $\overline{\Lambda}$ is indeed
about 0.05 $\sim$ 0.20 GeV) but $\lambda$ is of the order 1 GeV.
Meanwhile, for confining interaction, we shall only concentrate in
the region $q^+ = (y-y')K^+ \rightarrow 0$.  Thus,
\begin{equation}
	x_{\lambda ik} = {|\Delta K^-_{ik}| \over \lambda^2/K^+ 
		- \Delta K^-_{ik} } \simeq {(\kappa_\bot - \kappa'_\bot)^2
		\over (\kappa_\bot - \kappa'_\bot)^2 +(y-y')\lambda^2 }
		= {q_\bot^2 \over q_\bot^2 + \lambda^2 q^+/K^+},
\end{equation}
where $q_\bot$ and $q^+$ are the transverse and longitudinal momentum
carried by the exchanged gluons (see after Eq.(\ref{conp})). According 
to the properties of $f_{\lambda ij}$,  
\begin{equation} \left\{ \begin{array}{l}
	q_\bot^2 \leq \lambda^2 q^+/2K^+ \rightarrow x_{\lambda ik}
		\leq {1\over 3} \rightarrow f_{\lambda ik} = 1 \\  \\ 
	  \lambda^2 q^+/2K^+ < q_\bot^2 < 2\lambda^2 q^+/K^+ 
		\rightarrow {1\over 3} < x_{\lambda ik} < {2 \over 3} 
		\rightarrow f_{\lambda ik} ~{\rm smoothly~goes~from}~ 1 
		~{\rm to}~ 0 \\  \\
	 q_\bot^2 \geq 2\lambda^2 q^+/K^+ \rightarrow x_{\lambda ik}
		\geq {2\over 3} \rightarrow f_{\lambda ik} = 0
	\end{array}   \right.  .
\end{equation}  
Then the dominant contribution of the Fourier transformation of
$V_{Q\overline{Q} ij}$ for $q^+ \rightarrow 0$ is given by
\begin{eqnarray}
	 \int {dq^+ d^2q_\bot \over (2\pi)^2} && e^{i(q^+ x^- 
		+ q_\bot \cdot x_\bot)} \Bigg\{-{4g_\lambda^2 
		\over K^{+2}}~{1 \over (y-y')^2} f^2_{\lambda ik}
		\Bigg\}  \nonumber \\
	&& = - {g_\lambda^2 \over 2\pi^2} \int dq^+ e^{iq^+ x^-}
		{q^2_{\bot 1} \over q^{+2}}~{2J_1(|x_\bot|q_{\bot 1}) 
		\over |x_\bot| q_{\bot 1}} + \cdots , \label{conpg}
\end{eqnarray}
where $q_{\bot 1}^2 = \lambda^2 q^+ / 2K^+$, the dots denote the 
contribution from the region of ${1\over 3} < x_{\lambda ik} < 
{2 \over 3}$. Introducing a smooth function in this region is 
motivated for an easy control of possible singularities in the
further numerical computations \cite{Wilson94}. However, we 
found that the use of (\ref{sm2}) for $f_{\lambda ik}$ 
does not cause any new singularity for heavy quarkonium systems.
It is more interesting to see that the dominant contribution  of
$V_{Q\overline{Q} ij}$ (the first term in eq.(\ref{conpg})) has 
the same form of eq.(\ref{conp}) and results in the same
confining interactions of eqs.(\ref{lconp}) and (\ref{tconp})
(except for a factor $1/\sqrt{2}=f_s)$ that are obtained by using
eq.(\ref{sm2}). This shows that the 
existence of confining interactions in SRG is a general
feature in our formulation. The choice of eq.(\ref{sm2}) is just a 
simple way to introduce a convenient low-energy renormalization 
scale to nonperturbative QCD that makes the confining interaction 
explicit for heavy quarkonium systems.  Indeed, Perry and Wilson
have used different ways to show the existence of such a confining 
interactions \cite{Perry94,Wilson94a}.  The crucial point is that
{\it the above confining picture is indeed model independent since the
parameter $\lambda$ is a renormalization scale. Under the
consideration of SRG (see Section V), physical observables
are independent of $\lambda$. In other words, the confining
interactions obtained here are derived from QCD without introducing
new free parameters in SRG scheme.}

\subsection{Extension to heavy-light quark systems}

The heavy-light quark system (heavy hadrons containing 
one heavy quark) is  one of the most interesting 
topics in the current study of heavy hadron physics.
We now apply SRG to such systems.

The bare cutoff Hamiltonian we begin with for heavy-light 
quark systems is the combination of the heavy 
quark effective Hamiltonian (\ref{hqbeh}) and the 
full Hamiltonian for the light quarks and gluons \cite{Zhang93b}.  
We may also introduce
the residual center mass momentum for heavy-light systems,
$K^+ = \overline{\Lambda} v^+ = p^+_1 + k^+_1 = p^+_2 + k^+_2$,
$K_\bot = \overline{\Lambda} v_\bot = p_{1\bot} + k_{1\bot}
 = p_{2\bot} + k_{2\bot}$, 
where $\overline{\Lambda} = M_H-m_Q$, $p_1$ and $p_2$ are 
the light antiquark momenta and $k_1$ and $k_2$ the residual
momenta of the heavy quarks in the initial and final $Q\overline{q}$
states respectively.  

Following the general procedure, it is easy to find the 
nonperturbative part of the effective Hamiltonian for 
heavy-light quark systems,
\begin{equation}
	H_{\lambda 0 ij} = \theta({\lambda^2 \over K^+} -|\Delta 
		K_{ij}^-| ) \Big\{ H_{Q\overline{q}free ij}  
		+ V_{Q\overline{q} ij} \Big\} , \label{Qqeh}
\end{equation}
where
\begin{eqnarray}
	H_{Q\overline{q}free ij} &&= [2(2\pi)^3]^2 \delta^3
		(\bar{k}_1-\bar{k}_2) \delta^3(\bar{p}_1-\bar{p}_2) 
	 	\delta_{\lambda_1 \lambda_3} \delta_{\lambda_2 
		\lambda_4}	 \nonumber \\
	& & ~~~~~\times \Bigg\{ (y-1) \overline{\Lambda}^2 +
		{\kappa^2_{\bot} + m_q^2 \over y } 
		- {g^2 \over 2\pi^2 } C_f {\lambda^2 \over K^+}
		\ln \epsilon \Bigg\} , \label{Qqeh1} \\
	V_{Q\overline{q} ij}(y-y', && \kappa_\bot - \kappa'_\bot)=2(2\pi)^3 
		\delta^3(\bar{k}_1 + \bar{p}_1 - \bar{k}_2 - \bar{p}_2) 
		\delta_{\lambda_1 \lambda_3} \delta_{\lambda_2 \lambda_4} 
		{- 2g^2 (T^a)(T^a) \over (K^+)^2}	\nonumber \\ 
	&&~~~ \times  \Bigg\{ {2\over (y-y')^2}
		- \Bigg[ 2{(\kappa_\bot-\kappa'_\bot)^2 \over 
		(y - y')^2} - {\kappa^2_\bot -\kappa_\bot \cdot 
		\kappa'_\bot \over y(y-y') } - {\kappa_\bot \cdot
		\kappa'_\bot - (\kappa')^2_\bot \over y'(y-y')} 
		\Bigg] \nonumber \\
	& & ~~~~ \times \Bigg[ {\theta(B - \lambda^2) 
		\theta(B - A) \over (\kappa_\bot-\kappa'_\bot)^2 
		-(y-y')( {\kappa_\bot^2 \over y} - {(\kappa')_\bot^2
		\over y'})} + {\theta(A - \lambda^2) \theta(A - B) \over
		(\kappa_\bot-\kappa'_\bot)^2 + (y-y')^2 \overline{
		\Lambda}^2} \Bigg] \Bigg\}, \label{Qqvv}
\end{eqnarray} 
with  $B \equiv \Bigg|{(\kappa_\bot-\kappa'_\bot)^2 \over y-y'}
 - {\kappa_\bot^2 \over y} + {(\kappa')_\bot^2 \over y'} \Bigg| $
and the function $A$ has the same form as in quarkonium case.  Here 
we have also introduced $ y = p^+_1 / K^+ $, $\kappa_{\bot} = p_{1\bot} 
- y K_{\bot}$, but the range of $y$ is now given by $ 0 < y 
= {M_H \over \overline{\Lambda}} {p_1^+ \over P^+} < \infty $.

The heavy-light quark effective Hamiltonian is $m_Q$-independent. 
This is because in heavy-light quark systems the heavy quark 
kinetic energy can be treated as a perturbative correction to 
$H_{\lambda 0}$. Obviously the above $H_{\lambda 0}$ 
has the heavy quark spin and flavour symmetry.  Compared 
to the $V_{Q\overline{Q}}$, $V_{Q\overline{q}}$ interactions 
are much more complicated. But it is not difficult to check that 
the above $V_{Q\overline{q}}$ contains a confining interaction
when $q^+$ is small.  
The confining mechanism is the same for $Q\overline{Q}$ and 
$Q\overline{q}$ systems, as well as for $q\overline{q}$
systems.

In conclusion, we have obtained in this section the nonperturbative 
part of a confining QCD Hamiltonian for heavy-heavy and heavy-light 
 quark systems in the WCT scheme.  We have also shown that the confining 
mechanism is independent of a specific choice of the smearing 
functions. It is indeed also independent of the HQET, namely the
same confining picture occurs when we include all the $1/m_Q$ corrections
(see the discussion in the previous section).  We are now 
ready to study hadron states on the light-front and to explore how 
the WCT scheme works in the present formulation

\section{The WCT to heavy hadron bound states}

As we mentioned in the Introduction, 
the ideas of WCT to nonperturbative QCD
is to begin with the effective QCD Hamiltonian $H_\lambda = 
H_{\lambda 0} + H_{\lambda I}$. Then using the constituent picture
to solve nonperturbatively the hadronic bound state equations
governed by $H_{\lambda 0}$ and to determine the running coupling 
constant $g_\lambda$.  If one could properly choose the nonperturbative 
$H_{\lambda 0}$ such that $g_\lambda$ is arbitrarily 
small, then the corrections from $H_{\lambda I}$ could be computed 
perturbatively, and we would say that a WCT to nonperturbative 
QCD is realized. In this section, we shall study such a WCT to 
heavy hadron bound states.

\subsection{General structure of light-front bound state
equations}

In general, a hadronic bound state on the light-front can
be expanded in the Fock space composed of states with definite
number of particles \cite{Brodsky81,Zhang94a}. Formally, it
can be expressed as follows
\begin{equation}
        | \Psi(P^+, P_{\bot},\lambda_s) \rangle = \sum_{n,\lambda_i}
                \int \Big( \prod_i [d^3 \bar{p}] \Big)
                2 (2\pi)^3 \delta^3(\bar{P}-\sum_i \bar{p}_i)
                | n, \bar{p},\lambda_i \rangle
                \Phi_{n} (x_i,\kappa_{\bot i},\lambda_i), \label{lfwf}
\end{equation}
where $P^+, P_{\bot}$ are its total longitudinal and transverse 
momenta respectively and $\lambda_s$ its total helicity, $|n, 
\bar{p}, \lambda_i \rangle$ is a Fock state consisting of $n$
constituents, each of which carries momentum $\bar{p}_i$ and 
helicity $\lambda_i$ ($\sum_i \lambda_i = \lambda_s$); 
$\Phi(x_i,\kappa_{\bot i},\lambda_i)$ the corresponding amplitude
which only depends on the helicities $\lambda_i$, the longitudinal 
momentum fractions $x_i$, and the relative transverse momenta
$\kappa_{\bot i}$: $x_i = { p_i^+ \over P^+}$, $\kappa_{i\bot} 
= p_{i\bot} - x_i P_{\bot}.$

The eigenstate equation that the wave functions obey on the 
light-front is given by
\begin{equation}
        H_{LF} | P^+, P_{\bot},\lambda_s \rangle  = { P_{\bot}^2
                + M^2 \over P^+ } | P^+, P_{\bot},
                \lambda_s \rangle ,
\end{equation}
where $H_{LF}={P}^-$ the light-front Hamiltonian.  Explicitly, for 
a meson wave function, the corresponding light-front bound state 
equation is:
\begin{equation}
        \Big(M^2 - M_0^2 \Big)
                \left[\begin{array}{c} \Phi_{q\overline{q}} \\
                \Phi_{q\overline{q}g} \\ \vdots \end{array} \right]
                  = \left[ \begin{array}{ccc} \langle q \overline{q}
                | H_{int} | q \overline{q} \rangle & \langle q 
		\overline{q} | H_{int} | q \overline{q} g \rangle 
		& \cdots \\ \langle q \overline{q} g
                | H_{int} | q \overline{q} \rangle & \cdots & ~~  \\ \vdots &
                \ddots & ~~ \end{array} \right] \left[\begin{array}{c}
                \Phi_{q\overline{q}} \\ \Phi_{q\overline{q}g} \\ 
		\vdots \end{array} \right], \label{lfbe}
\end{equation}
where $M_0^2 = \sum_i { \kappa_{i\bot}^2 + m_i^2 \over x_i}$ the 
so-called invariant mass, $H_{int}$ the interaction part of 
$H_{LF}$.

Obviously, solving eq.(\ref{lfbe}) from QCD with the entire
Fock space is impossible.  A basic motivation of introducing
the WCT scheme is to simplify the complexities in solving 
the above equation. In the present framework, $H_{LF}= H_\lambda$,
where $H_\lambda$ has already decoupled from high energy states.  
Furthermore, the reseparation $H_\lambda=H_{\lambda 0} + 
H_{\lambda I}$ is another crucial step in WCT, where only
$H_{\lambda 0}$ is assumed to have the nonperturbative 
contribution to bound states through eq.(\ref{lfbe}), 
and $H_{\lambda I}$ is supposed to be a perturbative term 
which should not be considered when we try to solve
eq.(\ref{lfbe}) nonperturbatively.  

The next important step in the WCT scheme is the use of a 
constituent picture. The success of the constituent quark 
model suggests that we may only consider the valence 
quark Fock space in determining the hadronic bound 
states from $H_{\lambda 0}$.  In this picture, 
quarks and gluons must have constituent masses. 
This constituent picture can naturally be realized on
the light-front \cite{Wilson94}. However, an essential 
difference from the phenomenological constituent 
quark model description is that the constituent masses 
introduced here are $\lambda$ dependent.  The scale
dependence of constituent masses (as well as the effective
coupling constant) is determined by solving the bound 
states equation and fitting the physical quantities with 
experimental data. But for heavy quark mess, this 
$\lambda$-dependence can be ignored.
Once the constituent picture is introduced,
we can truncate the general expression of the light-front 
bound states to only including the valence quark Fock 
space.  The higher Fock space contributions can be 
recovered as a perturbative correction through $H_{\lambda I}$.  
Thus, eq.(\ref{lfwf}) for heavy quarkonia can be approximately
written as:
\begin{eqnarray}
        | \Psi (K^+, K_\bot, \lambda_s) \rangle = \sum_{\lambda_1 
		\lambda_2} \int [d^3\bar{k}_1] [d^3\bar{k}_2] && 
		2(2\pi)^3 \delta^3(\bar{K} - \bar{k}_1-\bar{k}_2) 
			\nonumber \\
		&& \times \phi_{Q\overline{Q}}
         	(y,\kappa_{\bot}) b^\dagger_v(k_1,\lambda_1) 
		d^\dagger_{-v}(k_2, \lambda_2) |0 \rangle ,   \label{QQwf}
\end{eqnarray}
where the wavefunction $\phi_{Q\overline{Q}}(y,\kappa_{\bot})$
may be mass dependent due to the kinetic energy in $H_{\lambda 0}$
[see (\ref{QQfe})] but it is spin independent in heavy mass limit. 
Also note that the heavy quarkonium states in heavy mass limit
are labelled by the residual center mass momentum $K^\mu$.
We may normalize eq.(\ref{QQwf}) as follows:
\begin{equation}
        \langle \Psi(K'^+,K'_\bot,\lambda_s') | \Psi(K^+,K_\bot,
		\lambda_s) \rangle = 2(2\pi)^3 K^+ \delta^3
		(\bar{K}-\bar{K}') \delta_{\lambda'_s \lambda_s}, 
		\label{nQQbs}
\end{equation}
which leads to 
\begin{equation}
       \int {dy d^2\kappa_{\bot} \over 2 (2\pi)^3} |\phi_{Q\overline{Q}}
		(y,\kappa_{\bot})|^2 = 1. \label{nQQwf}
\end{equation} 

With the above analysis on the quarkonium states, 
it is easy to derive the corresponding bound state equation.  
Let  $H_{LF} = H_{\lambda 0}$ of eq.(\ref{QQeh}), 
eq.(\ref{lfbe}) is reduced to 
\begin{eqnarray}
     \Bigg\{2\overline{\Lambda}^2 - {\overline{\Lambda}\over m_Q}
	\Big[2\kappa_\bot^2 + && \overline{\Lambda}^2(2y^2-2y +1)
	\Big] \Bigg\} \phi_{Q\overline{Q}}(y,k_{\bot}) 
	= \Bigg(-{g_\lambda^2\over 2 \pi^2} \lambda^2 C_f \ln\epsilon\Bigg)
		~\phi_{Q\overline{Q}}(y,k_{\bot}) \nonumber \\
	&& ~~ -4g_\lambda^2 (T^a)(T^a) \int {dy' d^2\kappa'_{\bot} \over 
		2(2\pi)^3} \Bigg\{ {1 \over (y-y')^2} \theta( 
		\lambda^2 - A) \nonumber \\ && ~~~~~~~~~~
	  +  {\overline{\Lambda}^2 \over (\kappa_{\bot} - 
	  \kappa_{\bot}')^2 + (y-y')^2 \overline{\Lambda}^2} \theta
		(A - \lambda^2) \Bigg\} \phi_{Q\overline{Q}}
		(y',\kappa'_{\bot}).  \label{QQbse}
\end{eqnarray}
This is the light-front bound state equation for heavy quarkonia 
in the WCT scheme. 

For the heavy mesons containing one heavy quark, similar consideration
leads to
\begin{eqnarray}
     \Bigg\{2\overline{\Lambda}^2 - \overline{\Lambda} \Big[ y 		
	\overline{\Lambda} - && {\kappa_\bot^2 + m^2_q(\lambda) 
	\over y\overline{\Lambda} } \Big] \Bigg\}\Phi_{Q\overline{q}}
	(y,k_{\bot},\lambda_1, \lambda_2) \nonumber \\
    && =  \Big(-{g_\lambda^2\over 2 \pi^2} \lambda^2 C_f \ln\epsilon\Big)
		~\Phi_{Q\overline{q}}(y,k_{\bot},
		\lambda_1, \lambda_2) \nonumber \\
	&& ~~~~ + (K^+)^2\int {dy' d^2\kappa'_{\bot} \over 
		2(2\pi)^3} V_{Q\overline{q}}(y-y',\kappa_{\bot}
		- \kappa_{\bot}')\Phi_{Q\overline{q}}
		(y',\kappa'_{\bot}, \lambda_1, \lambda_2),  
		\label{Qqbse}
\end{eqnarray}
where $V_{Q\overline{q}}$ is given by eq.(\ref{Qqvv}). Note 
that the light antiquark here is a brown muck, a current light 
antiquark surround by infinite gluons and $q\overline{q}$ 
pairs that results in a constituent quark mass $m_q$ which is
a function of $\lambda$.

\subsection{A general analysis of light-front wavefunctions}

A numerical computation to the bound state equations,
eqs.(\ref{QQbse}) and (\ref{Qqbse}), is actually not too difficult.
However, to have a deeper insight about the internal structure of
light-front bound states and to determine the scale dependence
of the effective coupling constant in $H_\lambda$, it is better to have
an analytic analysis.  For this propose, we would like to present
a general analysis of light-front hadronic wavefunctions and then 
use variational approach to solve the bound state equations.

The heavy hadronic wavefunctions in the heavy mass limit 
are rather simple. 
First of all, the heavy quark kinematics have already added 
some constraints on the general form of the light-front
wavefunction $\phi (x, \kappa_\bot)$.  When we introduce the 
residual longitudinal
momentum fraction $y$ for heavy quarks, the longitudinal 
momentum fraction dependence in $\phi$ is quite different 
for the heavy-heavy, heavy-light and light-light mesons.  

For the light-light mesons, such as pions, rhos, kaons etc., 
the wavefunction $\phi_{q\overline{q}}
(x,\kappa_\bot)$ must vanish at the endpoint $x=0$ or $1$.  
This can be seen from the kinetic energy term in 
eq.(\ref{lfbe}), where $M_0^2 = {\kappa_\bot^2 + m_1^2 \over x}
- {\kappa_\bot^2 + m_2^2 \over 1-x}$ for the valence Fock space.
To ensure that the bound state equation is well defined
in the entire range of momentum space, $|\phi_{q\overline{q}}
(x,\kappa_\bot)|^2$ must fall down to zero in the
longitudinal direction not slower than $1/x$
and $1/(1-x)$ when $x \rightarrow 0$ and $1$, respectively.
In other words, at least $\phi_{q\overline{q}}(x,\kappa_\bot)
 \sim \sqrt{x(1-x)}$ .
For heavy-light quark mesons, namely the $B$ and $D$ mesons, 
the wavefunction $\phi_{Q\overline{q}}(y,\kappa_\bot)$ is 
required to vanish at $y=0$, where $y$ is the residual 
longitudinal momentum fraction carried by the light quark. 
This is because the kinetic energy in eq.(\ref{Qqbse}) 
only contains a singularity at $y=0$.
On the other hand, since  $0 \leq y \leq \infty$, 
$\phi_{Q\overline{q}}(y,\kappa_\bot)$ should also vanish when 
$y \rightarrow \infty$.  Hence, a possible simple solution is
$\phi_{Q\overline{q}}(y,\kappa_\bot) \sim \sqrt{y}e^{-\alpha y}$ 
or $  \sqrt{y} e^{-\alpha y^2}$. Other form to suppress the 
singularity  at $y=0$, like $e^{-\alpha/y}$ instead of $\sqrt{y}$,
is also possible. For heavy quarkonia, 
$-\infty < y < \infty$, the normalization 
forces $\phi_{Q\overline{Q}}(y,\kappa_\bot)$ to vanish as
$y \rightarrow \pm \infty$. Thus, a simple solution may be 
$\phi_{Q\overline{Q}}(y,\kappa_\bot) \sim e^{-\alpha y^2}$.

On the other hand, the transverse momentum dependence in 
these light-front wavefunctions should be more or less
similar.  They all vanish at $\kappa_\bot 
\rightarrow \pm \infty$. A simple form of the $\kappa_\bot$
dependence for these wavefunctions is a Gaussian function:
$e^{- \kappa_\bot^2 / 2\omega^2}$. 

The above analysis of light-front wavefunctions is only 
based on the kinetic energy properties of the constituents.
Currently, many investigations on the hadronic
structures use phenomenological light-front wavefunctions. 
One of such wavefunctions that has been 
widely used in the study of heavy hadron structure is the
BSW wavefunction\cite{BSW},
\begin{equation}	\label{bsw}
	\phi_{BSW}(x,\kappa_\bot) = {\cal N} \sqrt{x(1-x)}
		~\exp\left(-{\kappa^2_\bot\over2\omega^2}\right)
		~\exp\left[-{M_H^2\over2\omega^2}(x-x_0)^2\right],
\end{equation}
where ${\cal N}$ is a normalization constant, $\omega$  
a parameter of order $\Lambda_{QCD}$, $x_0=({1\over2} 
-{m_1^2-m_2^2\over2M_H^2})$, and $M_H$, $m_1$, and $m_2$ are 
the hadron, quark, and antiquark masses respectively. In the 
heavy mass limit, the BSW wavefunction can be produced from 
our analysis based on the light-front bound state equations.

Explicitly, for heavy-light quark systems, such as the $B$ and 
$D$ mesons, one can easily find that in the heavy mass limit,
$m_1 = m_Q \sim M_H$, $m_q << m_Q$  so that $x_0 =0$ .
Meanwhile, we also have $M_H x = 
\overline{\Lambda} y$. Furthermore, the factor $\sqrt{x(1-x)}$ 
can be rewritten by $\sqrt{y}$ in according to the corresponding
bound state equation discussed above. Thus, the BSW wavefunction 
is reduced to
\begin{equation}	\label{Qqtwf}
	\phi_{Q\overline{q}}(y,\kappa_\bot) = {\cal N} \sqrt{y}
		~\exp\left(-{\kappa_\bot^2\over2\omega^2}\right)
	~\exp\left(-{\overline{\Lambda}^2\over2\omega^2}y^2\right).
\end{equation}
This agrees with our qualitative analysis given 
above. Indeed, using such a wavefunction we 
have already computed the universal Isgur-Wise function in 
$B \rightarrow D, D^*$ decays \cite{Cheung95}: $\xi( v \cdot v') 
= {1 \over v \cdot v'},$ and from which we obtained the slope 
of $\xi(v \cdot v')$ at the zero-recoil point, $\rho^2 = - \xi'(1) = 1$,
in excellent agreement with the recent CLCO result \cite{clco}
of $\rho^2 = 1.01 \pm 0.15 \pm 0.09$. 

For heavy quarkonia, such as the $b\overline{b}$ and $c\overline{c}$
states, $m_1 = m_2 = m_Q$ which leads to $x_0=1/2$ in eq.(\ref{bsw}). 
Also note that $M_H (x- 1/2) = \overline{\Lambda} y$, and the factor 
$\sqrt{x(1-x)}$ must be totally dropped as 
we have discussed form the quarkonium bound state equation. 
Then the BSW wavefunction for quarkonia is reduced to 
\begin{equation}	\label{QQtwf1}
	\phi_{Q\overline{Q}}(y,\kappa_\bot) = {\cal N} ~\exp \left(
		-{\kappa^2_\bot\over2\omega^2}\right) ~\exp\left(
		-{\overline{\Lambda}^2\over2\omega^2}y^2\right),
\end{equation}
which is the exact form as we expected from the qualitative 
analysis. Here we have not taken the limit of $m_Q 
\rightarrow \infty$ for heavy quarkonia. Thus a possible $m_Q$
dependence in wavefunction may be hidden in the parameter $\omega$.

In the following, we shall use the above Gaussian-type wavefunction 
ansatz to solve the light-front quarkonium bound state equation, 
and from which to determine qualitatively the nonperturbative 
scaling dynamics and to examine the ideas of WCT to heavy 
hadron bound states.  

\subsection{WCT of nonperturbative QCD description to heavy quarkonia}

We take the normalized wavefunction ansatz of (\ref{QQtwf1}),
\begin{equation}
	\phi_{Q\overline{Q}}(y,\kappa_\bot) = 4\sqrt{\overline{
		\Lambda}} \Bigg({\pi \over \omega_\lambda^2}\Bigg)^{3/4} 
		\exp \Bigg(-{\kappa_\bot^2 \over 2 \omega_\lambda^2} \Bigg) 
		\exp \Bigg( - { \overline{\Lambda}^2 \over 2 
		\omega_\lambda^2} y^2 \Bigg) ,	\label{QQtwf2}
\end{equation}
as a quarkonium trial wavefunction, where $\omega_\lambda$ means
that the wavefunction is also scale dependent. Substituting the 
above wavefunction into the quarkonium bound state
equation (\ref{QQbse}), we have 
\begin{equation}
	2\overline{\Lambda}^2 = {\cal E}_{kin} - {g_\lambda^2\over 
		2 \pi^2} \lambda^2 C_f \ln\epsilon + {\cal E}_{nonc} + 
		{\cal E}_{Coul} , \label{QQev}
\end{equation}
where ${\cal E}_{kin}={\overline{\Lambda}\over m_Q}\Big(3
\omega_\lambda^2 + \overline{\Lambda}^2\Big) $ represents 
the kinetic energy, ${\cal E}_{nonc}$ is the contribution 
of the noncancellation of the instantaneous interaction, 
\begin{equation}
{\cal E}_{nonc} = {g_\lambda^2\over 2\pi^2} C_f \lambda^2 \Bigg\{ 
	\gamma + \ln {\lambda^2 \epsilon \over 4\omega^2_\lambda}
		+  {\rm E}_1(\varpi^2) + {\sqrt{\pi}\over \varpi} 
		{\rm Erf}(\varpi) \Bigg\},  	\label{confe} 
\end{equation}
and ${\cal E}_{Coul}$ from the Coulomb interaction,
\begin{equation}
 {\cal E}_{Coul} = - {g_\lambda^2\over 2\pi^2} C_f 
		\lambda^2\Bigg\{ {\sqrt{\pi}\over \varpi} 
		\Big[1-~{\rm Erf}(\varpi)\Big] + {1 \over \varpi^2} 
		\Big[1 - e^{-\varpi^2} \Big] \Bigg\} ,
\end{equation}
where  $\gamma=0.57721566...$ the Euler constant, $\epsilon$ 
the small longitudinal momentum cutoff, the dimensionless 
$\varpi$ is defined by $\varpi = {\lambda^2 \over 2 \omega_\lambda 
\overline{\Lambda}}$, and E$_1$ and Erf are the exponential 
integral function and the error function, respectively.

We may rewrite the term $\ln {\lambda^2 \epsilon \over 4
\omega^2_\lambda}$ in ${\cal E}_{nonc}$ by $ \ln {\lambda^2 
\epsilon \over 4\omega^2_\lambda} = \ln\epsilon + \ln \varpi^2 
+ \ln {\overline{\Lambda}^2 \over \lambda^2}$ .
It shows that ${\cal E}_{nonc}$ contains a logarithmic 
divergence $\ln \epsilon$ which exactly cancels the same divergence 
from the self-energy correction, as expected, and the term 
$\ln \varpi^2$ is the logarithmic confining energy.

After the cancellation of the infrared $\ln \epsilon$ divergences 
in eq.(\ref{QQev}), the binding energy for heavy quarkonia is 
given by the kinetic energy plus the interaction energy:
\begin{equation}
	2\overline{\Lambda}^2 = {\cal E}_{kin} + {\cal E}_{conf}
		+ {\cal E}_{Coul} = {\overline{\Lambda} \over m_Q} 
		\Bigg\{3\omega_\lambda^2 + \overline{\Lambda}^2 
		\Bigg\} + {g_\lambda^2\over 2\pi^2}C_f \lambda^2 
		\Bigg\{F(\varpi)+\ln {\overline{\Lambda}^2 \over 
		\lambda^2} \Bigg\},  \label{sQQe}
\end{equation}
where
\begin{equation}	\label{dlsf}
	F(\varpi) = \gamma + \ln \varpi^2 + E_1(\varpi^2)
		- {\sqrt{\pi} \over \varpi}\Big[1 - 
		2{\rm Erf}(\varpi)\Big]-{1 \over \varpi^2} 
		\Big[1-e^{-\varpi^2}\Big] .
\end{equation}
In Fig.1, we plot the confining energy, the Coulomb energy and 
the totally interaction
energy as functions of $\varpi$ which is proportional
to the radial variable in the light-front space,
\begin{equation}
	\varpi \sim {1 \over \omega_\lambda} \sim r_l .
\end{equation} 
Fig.1 shows that {\it the total interaction energy is
a typical combination of the Coulomb interaction at short 
distance and a confining interaction at long distance 
that has been widely used in previous phenomenological 
description of hadronic states, but it is now analytically 
derived from QCD without introducing any free parameter
except for the QCD running coupling constant.}  
Furthermore, eq.(\ref{sQQe}) also indicates 
that without considering the kinetic energy, we cannot
find a stable quarkonium bound state.  The kinetic energy
balances the interaction energy and ensures the 
existence of a stable solution for (\ref{sQQe}).  Therefore,
the first order kinetic energy in HQET is an important 
nonperturbative effect in binding two heavy quarks, as
noticed first by Mannel et al.\cite{Mannel95}.

If we know the experimental value of the quarkonium
binding energy $\overline{\Lambda}$, minimizing eq.(\ref{sQQe}) 
can completely determine the parameter $\omega_\lambda$
and the coupling constant $g_\lambda$.  The precise 
value of quarkonium binding energy that can be compared with 
the data in Particle Data Group \cite{PDB} must include 
the spin-splitting energy ($1/m_Q$ corrections) which we 
will present in the forthcoming paper \cite{Zhang96}.
Here, to justify whether a WCT of nonperturbative QCD 
can become possible in the present formulation,
we will give a schematic calculation. It is known that 
$\overline{\Lambda}$ is of the same order  $\Lambda_{QCD}$
which is about $100 \sim 400$ MeV. We choose $\lambda$ as
a typical hadronic mass, $\lambda = 1$ GeV. The charmed 
and bottom quark masses used here are $m_c=1.4$ GeV and 
$m_b=4.8$ GeV. From the particle data \cite{PDB}, the
lowest charmonium states $M(\eta_c(1S))=2.98$ GeV and 
$M(J/\Psi(1S) = 3.10$ GeV, the bottomonium state
$M(\Upsilon(1S))=9.46$ GeV. Hence the binding energies 
of quarkonia $\overline{\Lambda} < 400$ MeV in the above
choice of heavy quark mass. To solve (\ref{sQQe}) we 
shall take several values of $\overline{\Lambda}$ within 
the above range. (In principle, it is not necessary to
must use a value of $\overline{\Lambda}$ below 400 MeV.)   
The results are 
listed in Tables I and II for charmonium and bottomonium, 
respectively, where $\omega_{\lambda 0}$ denotes the 
minimum point of the binding energy $\overline{\Lambda}$
(\ref{sQQe}). 

We see from the Tables I--II that the coupling 
constant $\alpha_\lambda = g^2_\lambda/4\pi$ is very small. 
For instance, with $\overline{\Lambda} = 200$ MeV,
we obtain
\begin{equation}
	\alpha_\lambda = \left\{ \begin{array}{ll} 0.02665~~~~~~~
		& {\rm charmonium,} \\
		0.06795~~~~ & {\rm bottomonium,} \end{array} \right.
\end{equation}
which is much smaller than that extrapolated from the 
running coupling constant in the naive perturbative 
QCD calculation. The parameter $\omega_{\lambda 0}$ 
is the mean value of the (transverse) momentum square of 
heavy quarks inside quarkonia:
\begin{equation}
	\langle k_\bot^2 \rangle = \omega_\lambda^2 .
\end{equation}
For charmonium, we can see that the resulting $\omega_\lambda$ 
are  typical values of $\Lambda_{QCD} \sim
\overline{\Lambda}$. The kinetic energy is about a half of the
interaction energy.  For bottomonium, we find that the binding 
energy $\overline{\Lambda}$ cannot be too large.  In fact, when
$\overline{\Lambda}$ is over about 260 MeV, eq.(\ref{sQQe}) 
has no solution.  Meanwhile, compared to charmonium, the 
effective coupling constant is relatively large 
(in contrast to the perturbative running coupling constant 
which is smaller with increasing $m_Q$
if it is taken as the energy scale).  Also the values of
$\omega_\lambda$ in bottomonium wavefunctions are larger 
than that in charmonium.  The difference between 
charmonium and bottomonium in the above nonperturbative 
calculation can be understood as follows.  As we know, 
in the nonrelativistic quark model, the quark momentum 
in quarkonia is proportional to the quark mass, 
$\omega_\lambda^2 \sim m_Q$ \cite{Gigg}.  Our relativistic 
QCD bound state solution exhibits such a property. This also 
indicates the first order kinetic energy has of the same order 
the leading Hamiltonian while all other terms are suppressed 
by the power of $1/m$ [see eq.(3.2)]. Therefore,
the kinetic energy must be treated nonperturbatively when
we apply HQET to heavy quarkonia. As a result, the bottomonium 
kinetic energy ($\sim \omega_\lambda^2$) becomes large as well.
To have a nonperturbative balance between the kinetic energy and
the interaction energy in the bound states, the coupling constant 
in bottomonium must be larger than that in charmonium. All these
properties now have been manifested in the solution of eq.(\ref{sQQe}). 
A more precise determination of $\alpha_\lambda$ (i.e., 
$g_\lambda$) requires an accurate computation of the low-lying 
quarkonium spectroscopy with the $1/m_Q$ corrections 
included\cite{Zhang96}.  Nevertheless, it has been shown that 
the effective coupling constant in $H_\lambda$ is very small 
at the hadronic mass scale.

In order to see how this weak coupling constant varies with the
scale $\lambda$, we take $\overline{\Lambda} =200$ MeV and
vary the value of $\lambda$ around 1 GeV. The result is listed 
in Table III. We find that the coupling constant is decreased
very faster with increasing $\lambda$.  In other words, {\it with 
a suitable choice of the hadronic mass scale $\lambda$
in SRG, we can make the
effective coupling constant $\alpha_\lambda$  in $H_\lambda$ 
arbitrarily small, and therefore the WCT of
nonperturbative QCD can be achieved in terms of $H_\lambda$ 
such that the corrections from $H_{\lambda I}$ can be truly
computed perturbatively}. 

\section{Discussion and Summary}

Thus far, the main ideas of the WCT to nonperturbative 
QCD proposed in the recent
publication \cite{Wilson94} have, at least qualitatively, 
been achieved for heavy quarkonia when $\lambda$ is around 
hadronic mass scale ($\sim$ 1 GeV). To have a deep
understanding of how WCT works to nonperturbative QCD, 
we shall study the $\lambda$-dependence of $\alpha_\lambda$
and discuss the physical implication of the WCT in this 
last section.

\subsection{Running effective coupling constant $\alpha_\lambda$
	in SRG}

From Tables I to III, we find that values of the dimensionless 
parameter $\varpi={\lambda^2\over 2\overline{\Lambda}
\omega_{\lambda 0}}$ are greater than 2.5. When $x > 2.5$, 
the exponential integral function and the error function 
are simply reduced to E$_1(x)=0$ and Erf$(x)=1$. Then,
eq.(\ref{sQQe}) is reduced to
\begin{equation}
	2\overline{\Lambda}^2 = {3\overline{\Lambda} \over m_Q} 	
		\omega_\lambda^2 - {g_\lambda^2\over 2\pi^2}C_f 
		\lambda^2 \Bigg\{{4\overline{\Lambda}^2 \omega_\lambda^2 
		\over \lambda^4} - \sqrt{\pi}~{2\overline{\Lambda}
		\omega_\lambda \over \lambda^2} - \Big(\gamma + \ln
		{\lambda^2 \over 4} -\ln \omega_\lambda^2\Big)~\Bigg\}
		+ {\overline{\Lambda}^3 \over m_Q},  \label{sQQe1}
\end{equation}
with an error less than $10^{-5}$.
Minimizing $\overline{\Lambda}$ with respect to $\omega_\lambda$, 
we obtain
\begin{equation}	\label{minimizing}
	 \Bigg\{ {3\overline{\Lambda} \over m_Q} - {g_\lambda^2 \over 
		2\pi^2}C_f{4\overline{\Lambda}^2 \over \lambda^2}
		\Bigg\} \omega_{\lambda}^2 = {g_\lambda^2 \over 
		2\pi^2}C_f \lambda^2 \Bigg\{ 1 - \sqrt{\pi}~
		{\overline{\Lambda} \omega_\lambda \over
		\lambda^2} \Bigg\}.
\end{equation}
Therefore, eq.(\ref{sQQe1}) becomes 
\begin{equation}
	2\overline{\Lambda}^2 = {g_\lambda^2\over 2\pi^2}C_f \lambda^2
		\Bigg\{ 1+ \gamma + \ln {\lambda^2 \over 4} - \ln
		\omega_{\lambda 0}^2 + \sqrt{\pi}~{\overline{\Lambda} 
		\omega_{\lambda 0}\over\lambda^2}~\Bigg\} + 
		{\overline{\Lambda}^3 \over m_Q},  \label{sQQe2}
\end{equation}
where $\omega_{\lambda 0}$ is a solution of (\ref{minimizing}). 
Since $\overline{\Lambda}$ is the binding energy of quarkonia,
it should be $\lambda$-independent. Thus, eqs.(\ref{minimizing}) 
and (\ref{sQQe2}) determine the $\lambda$-dependence of $g_\lambda$.  
The result is
\begin{eqnarray}
	\alpha_\lambda &=& {g^2_\lambda \over 4 \pi}
		= {\pi \over C_f}~ \Bigg( {\overline{\Lambda}^2
		\over \lambda^2}\Bigg)~{1\over a + b \ln {\lambda^2
		\over \overline{\Lambda}^2}},  \nonumber \\
	& \sim & {\overline{\Lambda}^2 \over \lambda^2} ~~~
		~~({\rm for~a~relatively~large}~ \lambda >> 
		\overline{\Lambda}), \label{running} 
\end{eqnarray}
where the coefficients $a$ and $b$ can be obtained by numerically 
solving eqs.(\ref{minimizing}) and (\ref{sQQe2}). The coefficient 
$b$ is almost a constant (with a weak dependence on $m_Q$ 
but independence on $\overline{\Lambda}$ and $\lambda$), while $a$
depends on both $\overline{\Lambda}$ and $m_Q$, and also slightly on
$\lambda$.  For $\lambda \geq 0.6$ GeV, the $\lambda$-dependence
in the parameter $a$ is negligible. In Fig.2, we plot the
$\lambda$-dependence of the effective coupling constant
$\alpha_\lambda$ for charmonium.  The dots are the numerical solutions 
of (\ref{sQQe}) and the solid line is given by the analytical
 form (\ref{running}) with $b=1.15$, and  $a=-0.25$ for
$\overline{\Lambda}=0.2$ GeV and $a=1.1$ for $\overline{\Lambda}
=0.4$ GeV. We can see that (\ref{running}) is a very good 
analytical solution of the eqs.(\ref{minimizing}) and 
(\ref{sQQe2}) [or of the minimizing eq.(\ref{sQQe})].

To understand the running behavior of $\alpha_\lambda$, we calculate
its SRG $\beta$-function. Denote the binding energy
\begin{equation}
	 \overline{\Lambda}=  \overline{\Lambda}
		(g_\lambda, \omega_\lambda, \lambda).
\end{equation}
The invariance of the binding energy $\overline{\Lambda}$ under
similarity renormalization group transformation means that 
 $\overline{\Lambda}$ determined from $H_{\lambda 0}$ and 
$H'_{\lambda' 0}$ must be the same for $\lambda \neq \lambda'$. 
Let $\lambda' = \lambda + \delta \lambda$, we obtain the 
corresponding similarity renormalization group equation
\begin{equation}	\label{srge}
	\Big( \lambda {\partial \over \partial \lambda} + \beta
		{\partial \over \partial g_\lambda} + \gamma_\omega
		{\partial \over \partial \omega_\lambda} \Big)
	\overline{\Lambda} (g_\lambda, \omega_\lambda, \lambda)=0,
\end{equation}
where the quantity $\beta$ is the similarity renormalization group
$\beta$ function which is defined by
\begin{equation}	\label{beta1}
	\beta(g_\lambda) = \lambda {d g_\lambda
		\over d\lambda}\Bigg|_{\lambda
		=\lambda(g_\lambda)},
\end{equation}
and $\gamma_\omega$ is an anomalous dimension that describes the
running properties of the bound state wavefunction.
The $\beta$ function can be determined by solving the SRG equation
(\ref{beta1}) or directly computed from eq.(\ref{running}),
\begin{eqnarray}
	\beta &=& - g_\lambda \Big(1 + {2b\over a + 2b \ln 
		{\lambda/\overline{\Lambda}}}\Big)\Big|_{\lambda
		=\lambda(g_\lambda)} \nonumber \\
	&\approx& - g_\lambda  ~~~~~( {\rm for~a~relatively~large}~~~
		\lambda >> \overline{\Lambda} ). \label{beta}
\end{eqnarray}
As we will see later, this $\beta$-function also indicates the 
existence of confining interaction in low energy region. 

\subsection{Why a WCT in low energy QCD region}

As we have pointed out in the last section we can always make the
effective coupling constant $\alpha_\lambda$ small with a suitable
choice of the low energy scale $\lambda$ such that a WCT to 
nonperturbative QCD can be realized on the light-front. The above 
dependence of $\alpha_\lambda$ with $\lambda$ shows that the 
effective coupling constant $\alpha_\lambda$ 
becomes smaller with larger $\lambda$. Thus, to have a WCT to 
nonperturbative QCD, $\lambda$ cannot be too small. In practice,
$\lambda$ is of the order 1 GeV. Now, the question is what is the
physical picture behind this dependence of $\alpha_\lambda$ 
with $\lambda$. 

As we discussed in the previous sections, the effective low 
energy Hamiltonian $H_{\lambda 0}$ is derived by integrating 
out the light-front high energy modes above $\lambda$ through 
the SRG. The existence of an explicit quark confining interaction 
in $H_{\lambda 0}$ (like to extract a mean-field in strongly 
correlated systems) is crucial to have a WCT 
in low-energy QCD region. Formally, as we have seen that the 
confining interaction in $H_{\lambda 0}$ comes from the 
noncancellation of instantaneous 
gluon exchange with energy below the scale $\lambda$.  With 
the larger $\lambda$, the more the instantaneous interaction 
contributes to $H_{\lambda 0}$ so that the confining interaction 
becomes stronger. A physical interpretation of the low 
energy scale $\lambda$, given by Wilson \cite{Wilson94a}, is 
that $\lambda$ corresponds to a up-bounded value of dynamical gluon 
mass in low-energy region.  Introducing $\lambda$ in SRG forces 
the gluon exchange energy in the $q\overline{q}$ effective 
interaction of $H_{\lambda 0}$ to be never less than $\lambda$. 
This is equivalent to say that dynamical gluons in low energy 
region have a mass of order $\lambda$ such that the energy of 
gluons propagating in $q\overline{q}$ channel cannot be smaller
than this dynamical mass.  A large $\lambda$ implies a large 
dynamical gluon mass which, on one hand, results in an explicit 
strong quark confining interaction in $H_{\lambda 0}$, and on 
the other hand, makes the residual quark-gluon interactions weak. 
The effective coupling constant $\alpha_\lambda$ characterizes 
these residual interactions ($H_{\lambda I}$ is expansed in 
terms of $\alpha_\lambda$). While, the smaller $\lambda$ implies to 
take the smaller dynamical gluon mass in low energy region. Then the 
explicit confining interactions in $H_{\lambda 0}$ is weakened 
and the residual interactions become stronger (i.e., some
of the confining interaction is hidden in the residual 
interactions). The limit $\lambda \rightarrow 0$ implies to keep
massless gluons in the low energy region. Then, our effective 
QCD theory turns back to the canonical form 
where the confining interaction is completely hidden in the 
beautiful but complicated quark-gluon interactions and the 
coupling constant becomes very strong in low energy region,
as we expected from the canonical theory. This is a physical 
interpretation of the above dependence of $\alpha_\lambda$ 
with $\lambda$. It also provides a physical picture why a
WCT can be applied to low energy QCD region for a finite
$\lambda$ in our description. 

To further examine the above understanding, we study the parameter 
$\omega_{\lambda 0}$ (the mean value of the transverse momentum
quarks carried inside the hadrons) as a function of $\lambda$.
The result is plotted in Fig. 3. From Fig. 3, we see that 
with increasing $\lambda$, $\omega_{\lambda 0}$ is decreased. 
Correspondingly, the distance between two quarks inside the 
quarkonia, $r_l \sim {1\over \omega_{\lambda 0}}$, is increased.  
This ensures that confining interaction becomes stronger with larger
$\lambda$ (as also shown in Table IV). Besides, Fig.2 also shows that 
$\lambda$ determined from the bound state equation can never 
go to zero. It implies that the dynamical gluons  in low energy 
region are most likely massive around a few hundreds MeV. 

It may also be worth noting that the low energy scale $\lambda$
introduced here is totally different from the  high energy scale 
$Q^2$ used in perturbative QCD.  The use of the scale $Q^2$
in perturbative QCD is to separate high and low energy region 
such that one should in principle integrate out the physics below 
the energy scale $Q^2$ [practically, instead of doing this 
(since none is able to do it), one usually uses factorization 
theorem to experimentally 
fit low energy physics] and then perturbatively compute the 
high energy physics above $Q^2$ from QCD. Here, the procedure 
seems to be an inverse processes of the perturbative QCD treatment.
We introduce the low-energy scale $\lambda$ and integrate out 
the high energy dynamics above $\lambda$. The left is just the
low energy QCD theory which describes the low energy physics.  
Normally, by integrating out high energy modes, the resulting 
low energy QCD is not neccessary to lead to the explicit quark 
confining interaction we obtained in $H_{\lambda 0}$. Therefore 
the theory is still nonperturbative. It is only on the 
light-front that the high energy modes contain long distance 
interactions which come from small longitudinal momentum 
of quarks and gluons [see the light-front dispersion relation 
$k^-={1\over k^+}(k_\bot^2+m^2)$, where a large light-front energy 
$k^-$ can be obtained from a small $k^+$]. Thus, integrating 
out light-front high energy modes results in an explicit quark
confining interactions in $H_{\lambda 0}$. Hence, the {\it light-front} 
formulation of the effective theory derived from SRG may be
another crucial point for making the WCT to nonperturbative QCD 
to be possible.
  
As we know in the canonical QCD theory, the confining interaction 
should become more important if the scale $Q^2$ would be smaller. 
Correspondingly the running coupling constant $\overline{\alpha}
(Q^2)$ becomes larger. Compare this fact with the above analysis, 
there may be an inverse correspondence between the effective 
coupling constant $\alpha_\lambda$ in $H_\lambda$ and the running 
coupling constant $\overline{\alpha} (Q^2)$ in the full QCD theory:
\begin{equation}	\label{invr}
	\alpha_\lambda \sim {1 \over \overline{\alpha}(Q^2)}~,
		~~~~ {\rm with}~~~~~  \lambda^2 \sim {1\over Q^2} .
\end{equation}
In other words, the weak-coupling treatment of the 
 confining Hamiltonian $H_\lambda$ may correspond to an 
inverse expansion of the original strong-coupling in the full QCD
theory. SRG just provides an implicative realization for such an 
expansion. 

More specifically, the running coupling constant in full QCD 
theory is given by
\begin{equation}
	 t= \int_{g_s}^{\overline{g}} {d g' \over \beta(g')},
\end{equation}
where $	t= {1\over 2}\ln {Q^2 \over \mu^2}$, and $Q^2$ is a
space-like momentum (the same as $\lambda^2$). Since the
similarity renormalization group $\beta$ function of eq.(\ref{beta})
is determined in the physical sector of low energy QCD dynamics, 
the low energy $\beta$ function of the running coupling constant 
$\overline{g}(Q^2)$ in the full theory should behave qualitatively 
the same.  Then we may take the $\beta(g)$ function in the above 
equation the same form as eq.(\ref{beta}) for $Q^2 << \Lambda^2_{QCD}$
(in low energy region). This leads to 
\begin{equation}	\label{sQ2}
	\overline{\alpha} (Q^2) = \alpha_s (\mu^2)
	{\mu^2 \over Q^2 } \equiv c_0 {\Lambda_{QCD}^2 \over Q^2}, ~~~
		~Q^2 << \Lambda^2_{QCD},
\end{equation}
where $c_0= \alpha_s (\mu^2)\mu^2/\Lambda_{QCD}^2$. This is 
consistent with eq.(\ref{running}) via eq.(\ref{invr}). To give 
a qualitative determination of the coefficient $c_0$ in (\ref{sQ2}),
we consider $\lambda =0.75 \sim 1.5$ GeV and $\overline{\Lambda}=0.2$
GeV, then $\lambda^2 = (14 \sim 56) \overline{\Lambda}^2 >>
\overline{\Lambda}^2$. From (\ref{running}), we have:
\begin{equation}
	\alpha_\lambda = (1.0 \sim 1.5) ~{\overline{\Lambda}^2
		\over \lambda^2}.
\end{equation}
The corresponding $Q^2 \sim {1\over \lambda^2} <<
\overline{\Lambda}^2 \sim \Lambda_{QCD}^2$. From 
(\ref{invr}), it follows that for $Q^2 << \Lambda_{QCD}^2$ the 
running coupling constant in the full QCD theory may be 
approximately given by 
\begin{equation}	\label{sQ21}
	\overline{\alpha}(Q^2) = (1.0 \sim 1.5)
		{\Lambda_{QCD}^2 \over Q^2}.
\end{equation}
This is just a qualitative estimation of the running coupling 
constant in the full QCD theory in low energy region that is 
obtained by the use of the same $\beta$-function determined 
in SRG from our effective low energy QCD description. 

Up to date, no one precisely knows how the full QCD coupling
constant $\alpha_s$ varies in low energy region.  However,
it is interesting to see that the running coupling constant 
given by eq.(\ref{sQ2}) for small $Q^2$ is indeed the 
basic assumption of the Richardson $Q\overline{Q}$ potential 
\cite{Rich79}:
\begin{equation}	\label{QQp}
	V(Q^2) = - C_f {\overline{\alpha}(Q^2) \over Q^2} ,
\end{equation}
where
\begin{equation}
	\overline{\alpha}(Q^2) = {12 \pi \over (33-2N_f)
		\ln(1 + Q^2/\Lambda_{QCD}^2)} .
\end{equation}
The Richardson $Q\overline{Q}$ potential is proposed to
exhibit the asymptotic freedom of QCD in short distance and
a confining potential in large distance limit.  From eq.(\ref{QQp}),
we see that for small $Q^2 ~(Q^2 << \Lambda_{QCD}^2) $,
\begin{equation}	\label{sQ22}
	\overline{\alpha}(Q^2)  \sim {12 \pi \over
		33-2N_f} ~ {\Lambda_{QCD}^2 \over Q^2},
\end{equation}
Comparing with eqs.(\ref{sQ2}) and (\ref{sQ22}), we have from 
the Richardson $Q\overline{Q}$ potential(with $N_f=3$ 
\cite{Rich79})
\begin{eqnarray}	\label{last}
	c_0 = {12 \pi \over (33-2N_f)}  =  1.4~.
\end{eqnarray}
This result agrees very well with eq.(\ref{sQ21}). Hence,
the dependence of $\alpha_\lambda$ with $\lambda$ determined 
in our description is essentially a consequence of exhibiting 
an explicit quark confining interaction in our low energy 
QCD Hamiltonian.  The possible inverse relation between 
$\alpha_\lambda$ with $\alpha_s(Q^2)$ discussed above can be
considered as another reason why the present effective theory 
to low energy of QCD can be treated as a weak-couple problem.

Finally, we should also note that there is an very interesting 
question remained to be answered: where is the cross-point (or 
a cross region) between low energy and high energy domains that 
let $\alpha_\lambda$ and $\alpha_s$ match at this point (or region). 
At this point, we have no answer for it but this is a very
important open question for a complete understanding of QCD
dynamics \cite{Wilson94}.

\subsection{A possible criterion for the $\lambda$ scale-fixing}

Because of the dependence of $\alpha_\lambda$ with $\lambda$
(\ref{running}), the realization of WCT to nonperturbative QCD
in our description depends on a suitable choice of $\lambda$,
as we have mentioned. One may further ask what is a suitable 
choice of $\lambda$ so that we can always treat low energy QCD 
as a weak-coupling problem, and whether there is any ambiguous 
in the choice of $\lambda$.

Table IV tells us that the confining interaction
plays a more important role than the Coulomb interaction in the
determination of the quarkonium bound states on the light-front.
This result is different from the usual understanding in the
nonrelativistic phenomenological description that the
dominant contribution in heavy quarkonium spectroscopy
is the Coulomb interaction.  This discrepancy can be
understood as follows.  The currently relativistic
light-front description for heavy quark systems mostly
uses the heavy quark masses of $m_c=1.3 \sim 1.4$ GeV
and $m_b=4.7 \sim 4.8$ GeV or less. In Particle Data Group
\cite{PDB}, $m_b=1.0 \sim 1.6$ GeV and $m_c=4.1 \sim 4.5$
GeV. Thus, the heavy quarkonium binding energies,
$\overline{\Lambda}= M_H - 2m_Q$, might be positive
[the lowest charmonium ground state $M(\eta_c(1S))=2.98$  
GeV, and the bottomonium $M(\Upsilon(1S)) = 9.46$ GeV].
Therefore, the Coulomb energy becomes not important for
a relativistic description in quarkonia. The
dominant contribution for binding quarkonium states must come
from the nonperturbative balance between the kinetic energy
and the confining energy. While, in the nonrelativistic
phenomenological description, one used larger quark masses,
$m_c > 1.8$ GeV and $m_b > 5.1$ GeV \cite{IGWS}, such that
the binding energy is forced to be negative.  As a result,
the Coulomb interactions must be dominant in this picture.
Of course, on the light-front, the structure of the bound
state equation is different from the nonrelativistic
Schr\"{o}dinger equation.  There is no direct comparison.
A real solution to the above discrepancy can be obtained
after including the spin-splitting interactions
($1/m_Q$-corrections). 

Nevertheless, from the above analysis, we have seen that 
the quark confining energy is a dominant contribution to 
the binding energy of bound states. A large binding energy 
$\overline{\Lambda}$ requires a large contribution from
the confining interaction and therefore a large value of $\lambda$ 
is needed. A small $\overline{\Lambda}$ requires a relatively 
small effect from the confining interaction so that a small 
$\lambda$ can be chosen. In all the cases, the effective 
coupling constant $\alpha_\lambda$ is kept at a small value. 
Thus, by choosing $\lambda$ to fix the rate $\overline{\Lambda}
\over \lambda$ at a certain value, the resulting $\alpha_\lambda$ 
can remain to be small. The WCT to low energy QCD dynamics is 
then always applicable. This is a criterion for fixing $\lambda$ 
in our description, similar to the scale-fixing procedure of 
Brodsky-Lepage-Mackenzie in solving scale ambiguity for the 
usual perturbation theory\cite{BLM}.
Of course, in principle one can choose an arbitrary value of
$\lambda$, physics should not be changed since the dependence
of $\alpha_\lambda$ with $\lambda$ is determined by the SRG 
invariance.  The above criterion for the scale-fixing is useful
for the realization of a WCT to perturbative QCD in our 
description.

In practice, we also find that there is no much freedom in the
choice of $\lambda$ value. The scale dependence of 
wavefunctions also provides 
a restriction on the range of $\lambda$.  For quarkonia, 
$\omega_\lambda$ is decreased with increasing $\lambda$. 
However, $\omega_\lambda$ is proportional to the mean value of
the (transverse) momentum square of quarks inside 
quarkonia, which characterizes the size of hadrons. 
Therefore $\lambda$ should not be too large for the best 
description of bound states. 

On the other hand, as we argued that gluons are
massive at the hadronic scale.  Hence, the lowest 
bound value of $\lambda$ can be considered
as the constituent gluon mass. As an example, one may take 
$\lambda$ to be a constituent gluon mass (about a half 
of the glueball masses, such as the recent possible evidences 
of $f_0(1500)$ and $\xi(2230)$ \cite{Glueball}), 
\begin{equation}	\label{lam}
	\lambda :~ (0.75 \sim 1.5) ~{\rm GeV}.
\end{equation}
Then the effective coupling constant (with $\overline{\Lambda}
= 200$ MeV) is
\begin{equation}
	\alpha_\lambda = 0.06 \sim 0.01 
\end{equation}
which is very small.  Only a finite $\lambda$ at the order
of hadronic mass scale can effectively turn the nonperturbative 
contribution in the higher order processes into the long distance 
two-body confining interactions through SRG.  For the range of 
eq.(\ref{lam}), we further have, 
\begin{equation}
 	\omega_{\lambda 0} = 0.24 \sim 0.2 ~{\rm GeV},	\label{omm}
\end{equation}
which leads to
\begin{equation}
	\langle r \rangle \sim {1 \over \omega_{\lambda 0}}
		= 0.8 \sim 1.0 ~{\rm fm} .
\end{equation}
This gives a resonable quarkonium size. Thus, we have provided
a qualitative criterion and a quantitative analysis on the choice
of $\lambda$ such that a true WCT to nonperturbative QCD can 
always be realized by $H_\lambda$.

\subsection{Summary}
 
In summary, to examine the 
weak-coupling treatment of nonperturbative QCD recently 
proposed in Ref. \cite{Wilson94}, we have studied explicitly 
the heavy quark bound state problem, based on the light-front 
heavy quark effective theory of QCD \cite{Zhang95,Cheung95}.
Firstly, we have used the similarity renormalization group approach
\cite{Wilson94,Glazek94} to derive the effective confining
Hamiltonian in the low energy scale for heavy quarks in heavy
mass limit. To make the similarity renormalization
approach practically manable, we have introduced a local
cutoff scheme (\ref{ctf}) to the bare QCD (and the effective
heavy quark) Hamiltonian, which simplifies the cutoff
scheme in \cite{Wilson94}.  Meanwhile we have also introduced
a simple smearing function $f_{\lambda ij}$ (\ref{sm2})
to the similarity renormalization group approach which
further simplifies the original formulation of
\cite{Wilson94}. The resulting low-energy effective QCD
Hamiltonian of heavy quark interactions exhibits the coexistence
of a confining interaction and a Coulomb interaction on the
light-front without introducing any additional free parameter
except for the effective QCD coupling constant.

The weak-coupling treatment can be realized for nonperturbative
QCD because the light-front similarity renormalization group 
approach with a finite $\lambda$ extracts an explicit quark 
confining interaction from the higher order nontrivial quark-gluon 
interactions into $H_{\lambda 0}$ so that the residual quark-gluon 
interactions become weak even in low energy region.
The weak-coupling treatment of nonperturbative
QCD is manable for $\lambda$ being around the hadronic
mass scale.  The similarity renormalization group invariance
can remove the $\lambda$-dependence in all the physical
observables obtained from the effective $H_\lambda$. 
The well-defined bound state description in QED is a special 
case ($\lambda \rightarrow 0$) of the SRG approach.  The whole 
idea of the WCT to nonperturbative QCD is originally motivated 
from the bound state description of QED \cite{Wilson94}.
Now, a consistent connection between QCD and QED and
their differences in determining bound states is explicitly
examined on the light-front.
 
The applications of the present theory to heavy quarkonium 
spectroscopy and various heavy quarkonium annihilation
and production processes 
can be simply achieved by numerically solving the bound 
state equations (\ref{QQbse}), and
by further including the $1/m_Q$ corrections (which naturally
leads to the spin splitting interactions).  The extension 
of the computations to heavy-light quark systems is 
straightforward. The extension of the 
present work to light-light hadrons requires the
 understanding of chiral symmetry breaking 
in QCD which is a new challenge to nonperturbative QCD 
on the light-front.  Nevertheless, the present work has 
provided a preliminary realization to the weak-coupling 
treatment of nonperturbative QCD proposed recently by
Wilson et al. \cite{Wilson94}. The new research 
along this direction is in progress.

\acknowledgements
The author would like to thank H. Y. Cheng, C. Y. Cheung,
A. Harindranath, R. J. Perry, K. G. Wilson, and T. M. Yan for 
many fruitful discussions, and specially H. Y. Cheng for his 
carefully reading the manuscript. This work is supported in 
part by National Science Council under Grant Nos. NSC 
84-2816-M-001-012L and NSC 86-2816-M001-009R-L.  An earlier
and longer version of this paper can be found: hep-ph/9510428.

\newpage


\newpage


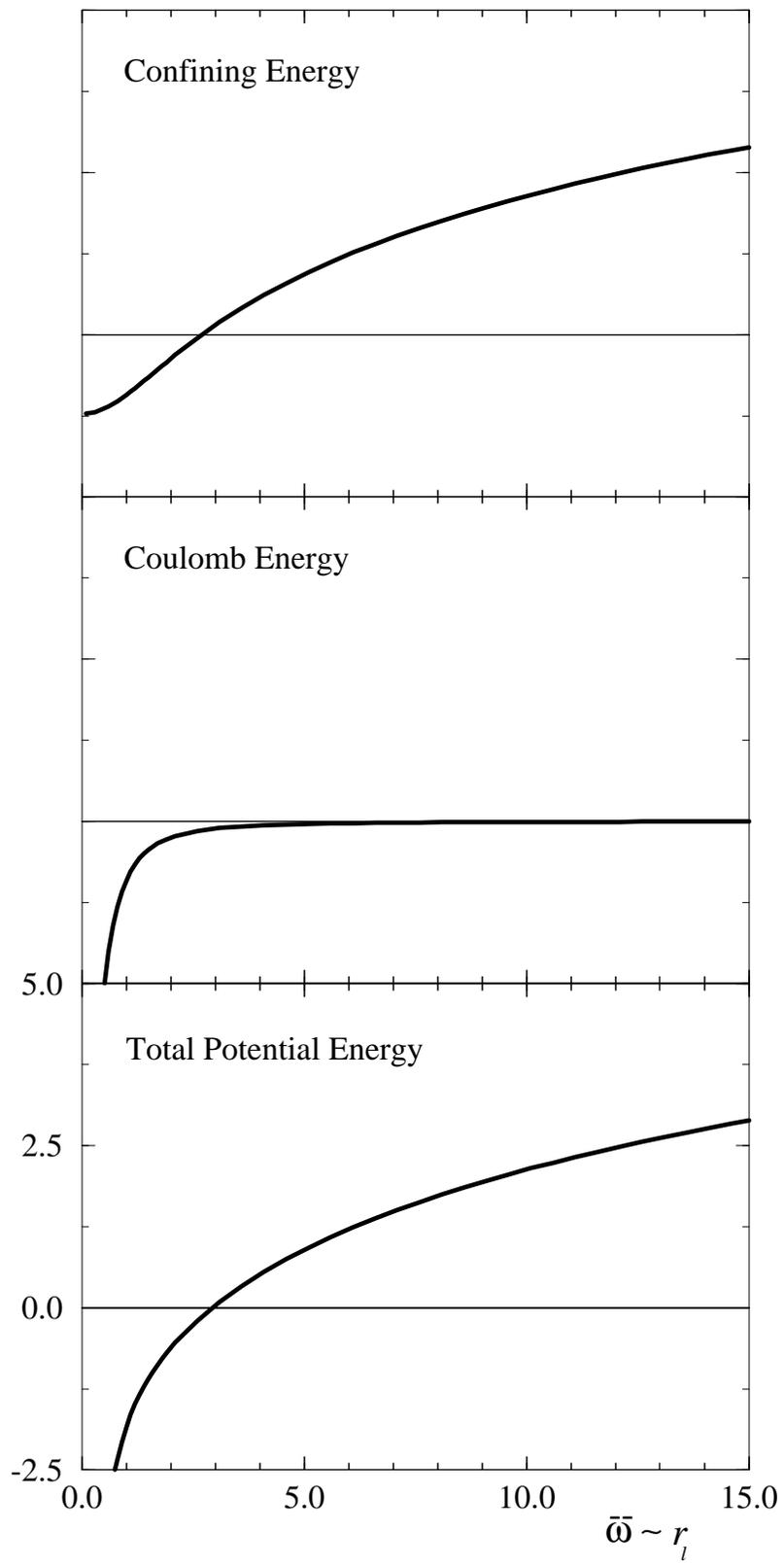
\begin{figure}
\caption[]{A plot of the confining energy,
the Coulomb energy and the total interaction 
energy as functions of the dimensionless variable 
$\varpi$ but $\varpi$ is proportional to $\sim r_l$
via $\omega_\lambda$. The energies are scaled by the 
factor ${g^2_\lambda \lambda^2 \over 2 \pi^2} C_f$.}
\end{figure}

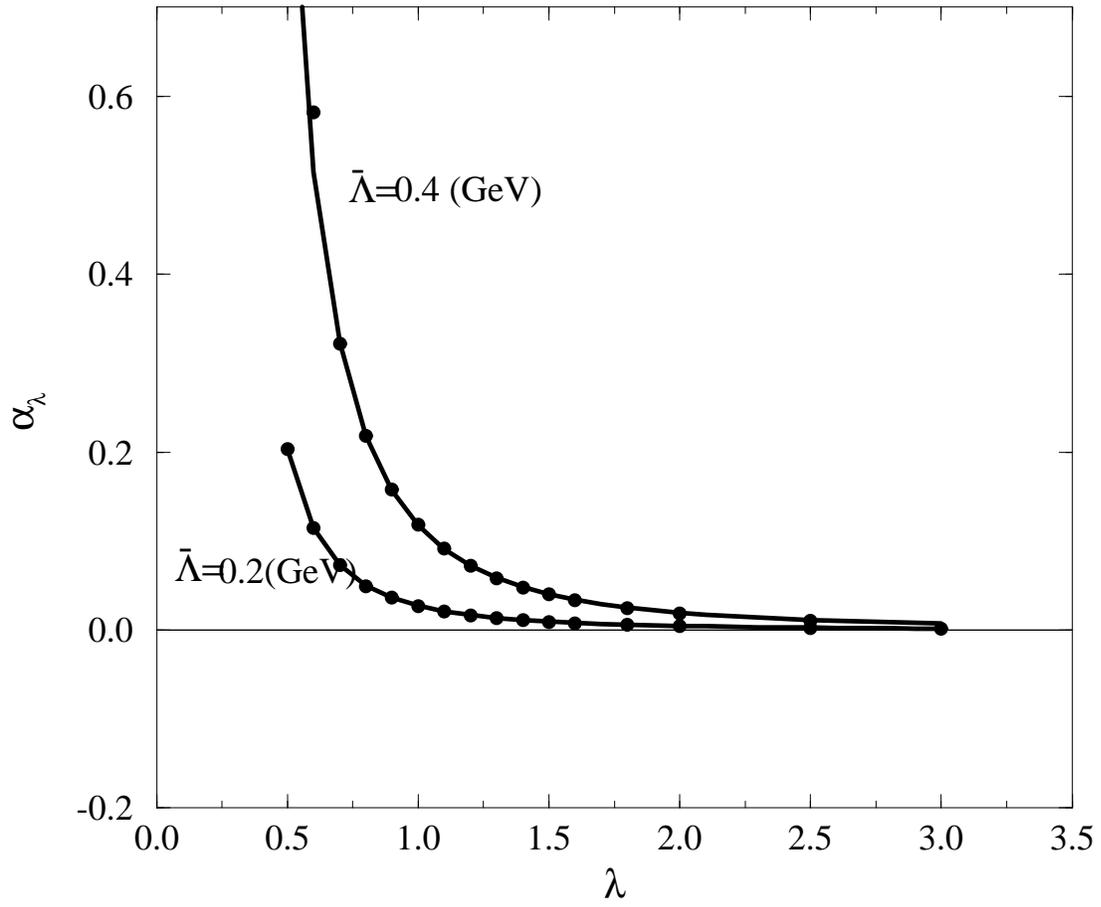
\begin{figure}
\caption[]{The $\lambda$-dependence of the effective
coupling constant $\alpha_\lambda$. The solid line 
is given by the analytical result (\ref{running}),
and the dots are obtained by numerically
minimizing the quarkonium binding energy 
(\ref{sQQe}). Here $\lambda$ is given in units of GeV.}
\end{figure}

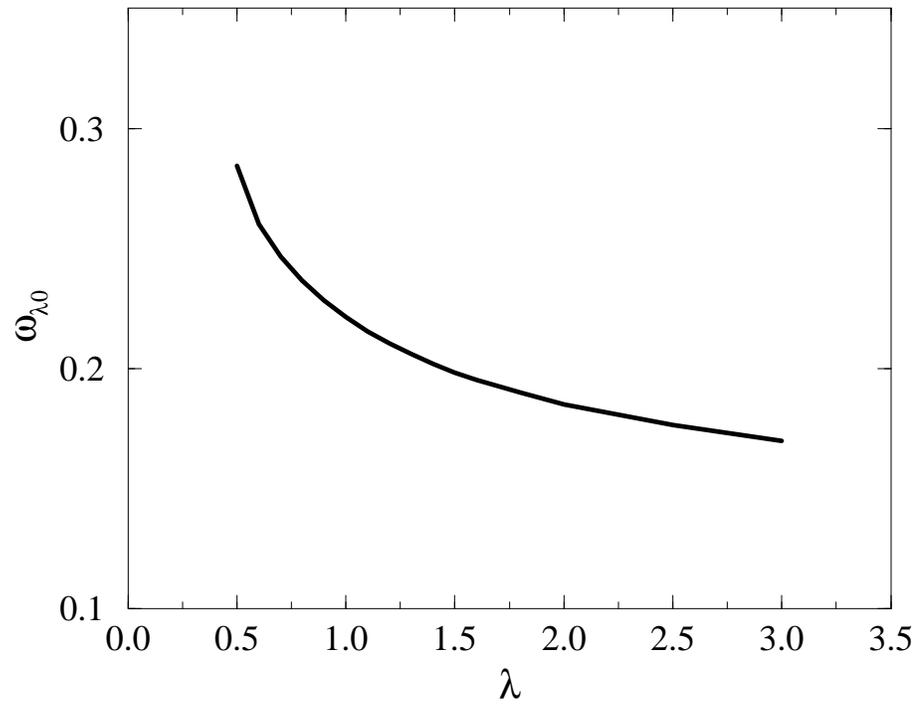
\begin{figure}
\caption[]{The $\lambda$-dependence of the wavefunction
parameter $\omega_{\lambda 0}$ which is the solution of
minimizing the quarkonium binding energy (\ref{sQQe}), 
where $\lambda$ is given in units of GeV.}
\end{figure}

\newpage

\begin{center} TABLES 

\vspace{0.5in}

\begin{tabular}{|c|c|c|c|c|c|}  \multicolumn{6}{c}{Table I. 
Solution for charmonium ground state with $m_c=1.4$ GeV} \\
\hline\hline $\overline{\Lambda}$ 
(GeV) &~~~~ $\alpha_\lambda$~~~~& $\omega_{\lambda 0}$
(GeV) & \multicolumn{3}{c|} {${\cal E}_{kin}$(GeV$^2$) + 
${\cal E}_{pot}$(GeV$^2$) = $2 \overline{\Lambda}^2$(GeV$^2$)} \\ \hline
0.2 & 0.02665 & 0.222 & ~~0.026836 ~~& ~~0.053173 ~~&~ 0.080009~ \\ \hline
0.3 & 0.06480 & 0.275 & 0.067902 & 0.112018 & 0.179920 \\ \hline
0.4 & 0.11831 & 0.314 & 0.130225 & 0.189781 & 0.320006 \\ \hline
\hline \end{tabular}

\vspace{0.5in}

\begin{tabular}{|c|c|c|c|c|c|}  \multicolumn{6}{c}{Table II. 
Solution for bottomonium ground state with $m_b=4.8$ GeV} \\ \hline\hline
$\overline{\Lambda}$ (GeV) & $\alpha_\lambda$ & $\omega_{\lambda 0}$
(GeV) & \multicolumn{3}{c|} {${\cal E}_{kin}$(GeV$^2$) + 
${\cal E}_{pot}$(GeV$^2$) = $2 \overline{\Lambda}^2$(GeV$^2$)} \\ \hline
0.15 & 0.029965 & 0.492 & ~~0.023397~~ &~~ 0.021602~~ &~ 0.044999~ \\ 
\hline 0.20 & 0.06795 & 0.623 & 0.050183 & 0.029816 & 0.079999 \\ \hline
0.25 & 0.1385 & 0.779 & 0.098074 & 0.026930 & 0.125004 \\ \hline
\hline \end{tabular}

\vspace{0.5in}

\begin{tabular}{|c|c|c|c|c|} \multicolumn{5}{c}{Table III. Some 
numerical solution on the $\lambda$-dependence} \\ 
\multicolumn{5}{c}{of the weak coupling constant $\alpha_\lambda$.}\\
\hline\hline ~~& \multicolumn{2}{c|}{charmonium} &
\multicolumn{2}{c|}{bottomonium}\\ \hline $\lambda$ (GeV) &~~~~~ 
$\alpha_\lambda$~~~~~& ~$\omega_{\lambda 0}$ (GeV)~&~~~~~
$\alpha_\lambda$~~~~~& $\omega_{\lambda 0}$ (GeV) \\ \hline
0.75 & 0.05960 & 0.241 & ~~~~ & ~~ ~~\\ \hline
1.0 & 0.02665 & 0.222 & 0.06795 & 0.623 \\ \hline
1.5 & 0.00912 & 0.199 & 0.01607 & 0.478 \\ \hline 
2.0 & 0.00441 & 0.185 & 0.00695 & 0.427 \\ \hline
\hline \end{tabular}

\vspace{0.5in}

\begin{tabular}{|c|c|c|c|c|c|c|c|c|} \multicolumn{9}{c}{Table IV. The $\lambda$-dependence of various interactions to the binding energy}\\ \hline\hline $\lambda$ (GeV) & ~~~0.5~~~ & ~~~0.75~~~ & ~~~1.0~~~ & ~~~1.2
~~~ & ~~~1.4~~~ & ~~~1.8~~~ & ~~~2.0~~~& ~~~3.0~~~ \\ \hline
${\cal E}_{kin}$ (GeV$^2$) & 0.04 & 0.031& 0.027& 0.025& 0.023& 
0.021& 0.02& 0.018\\ \hline
${\cal E}_{conf}$ (GeV$^2$) & 0.049 & 0.050& 0.053& 0.055& 
0.057& 0.059& 0.06 &0.062 \\ \hline
${\cal E}_{Col}$ (GeV$^2$)  & -0.009 & -0.001& -0.001& 0.0& 
0.0& 0.0 &0.0 & 0.0 \\ \hline \hline \end{tabular} \\ \ \\
\end{center}

\end{document}